\documentclass[aps,showpacs,twocolumn,preprintnumbers,amsmath,amssymb,pra]{revtex4-1}
\usepackage{graphicx}% Include figure files
\usepackage{dcolumn}% Align table columns on decimal point
\usepackage{bm}% bold math
\usepackage{color}
\begin{document}

\title{Baseband Modulation Instability as the Origin of Rogue Waves}

%\author{Fabio Baronio} \email{fabio.baronio@unibs.it} 
%\affiliation{Dipartimento di Ingegneria dell'Informazione, Universit\`a di Brescia, \\
%Via Branze 38, 25123 Brescia, Italy}
%\author{Matteo Conforti}
%\affiliation{PhLAM/IRCICA UMR 8523/USR 3380,  CNRS-Universit\'e Lille 1,  F-59655 Villeneuve d'Ascq, France}
%\author{Shihua Chen}
%\affiliation{Department of Physics, Southeast University, Nanjing 211189,China}
%\author{Philippe Grelu}
%\affiliation{Laboratoire Interdisciplinaire Carnot de Bourgogne, UMR 6303 CNRS-Universit\'e de Bourgogne, BP 47870 Dijon Cedex 21078, France}
%\author{Stefan Wabnitz}
%\affiliation{Dipartimento di Ingegneria dell'Informazione, Universit\`a di Brescia, \\
%Via Branze 38, 25123 Brescia, Italy}

\author{Fabio Baronio$^{1*}$, Shihua Chen$^2$, Philippe Grelu$^3$, Stefan Wabnitz$^1$, and Matteo Conforti$^4$} 
\affiliation{$^1$Dipartimento di Ingegneria dell'Informazione, Universit\`a di Brescia, Via Branze 38, 25123 Brescia, Italy,\\
$^2$Department of Physics, Southeast University, Nanjing 211189,China,\\
$^3$Laboratoire Interdisciplinaire Carnot de Bourgogne, UMR 6303 CNRS-Universit\'e de Bourgogne, BP 47870 Dijon Cedex 21078, France,\\
$^4$PhLAM/IRCICA UMR 8523/USR 3380,  CNRS-Universit\'e Lille 1,  F-59655 Villeneuve d'Ascq, France.}

\begin{abstract}
We study the existence and properties of rogue wave solutions 
in different nonlinear wave evolution models that are commonly used in optics and hydrodynamics.
In particular, we consider Fokas-Lenells
equation, the defocusing vector nolinear 
Schr\"odinger equation, and the long-wave-short-wave resonance equation. We 
show that rogue wave solutions in all of these models exist in the 
subset of parameters where modulation instability is present, if and only if the unstable sideband
spectrum also contains cw or zero-frequency perturbations as a limiting 
case (baseband instability). We numerically confirm that
rogue waves may only be excited from a weakly perturbed cw whenever the baseband instability is present. Conversely, modulation instability leads to nonlinear periodic oscillations. 
\end{abstract}

%\ocis{190.0190, 190.3100, 190.4223, 190.5530}% REPLACE WITH CORRECT OCIS CODES FOR YOUR ARTICLE
                          %% NOTE: \ocis{} IS ALIASED TO \pacs{} BUT MUST
                          %% FORMAT THE TERMS CORRECTLY FOR EACH JOURNAL
													
\pacs{05.45.Yv, 02.30.Ik, 42.65.Tg}

\maketitle

\section{Introduction}

Many nonlinear wave equations associated with different physical systems exhibit the emergence of extreme, high-amplitude
events that occur with low probability, and yet may have dramatic consequences. 

Perhaps the most widely known examples of such processes are the giant oceanic \textit{rogue waves} 
\cite{hopkin04} that unexpectedly grow with a great destructive power from the average sea level fluctuations. 
This makes the study of 
rogue waves a very important problem for ocean liners and hydrotechnic constructions \cite{perkin06,pelinosky08}. 
Hence, it is not surprising that the phenomenon of rogue waves has attracted the ample attention of 
oceanographers over the last decade.
Although the existence of rogue waves has been confirmed by multiple observations, 
uncertainty still remains on their fundamental origins  \cite{kharif09}. %This hampers systematic approaches to study their characteristics, including the predictability of their appearance.

In recent years, research on oceanic rogue waves has also drawn the interest of researchers 
in many other domains of physics and enginering applications, which share similar complexity features: in particular, consider nonlinear optics 
\cite{dud14}. The ongoing debate on the origin and definition of rogue waves has stimulated the comparison of their predictions and observations in 
hydrodynamics and optics, since analogous dynamics can be
identified on the basis of their common mathematical models \cite{onorato13}. 

So far, the focusing nonlinear Schr\"odinger equation (NLSE) has played a pivotal
role as universal model for rogue wave solutions, boh in optics and in hydrodynamics. For example, the Peregrine soliton, first
predicted as far as $30$ years ago \cite{peregrine83}, is the simplest rogue-wave solution 
of the focusing NLSE. This rogue wave has only recently been experimentally 
observed in optical fibers \cite{kibler10}, water-wave tanks \cite{amin11}, and plasmas \cite{bailung11}.

%Moving beyond the standard focusing NLSE description in order to model more general and important
%classes of physical systems is both relevant and necessary.
For several systems the standard focusing NLSE turns out to be an oversimplified description: this fact pushes the research to move beyond this model.
In this direction, recent developments consist in including the effect of dissipative terms. In fact, a substantial supply of energy
(f.i., from the wind in oceanography, or from a pumping source in laser cavities) is generally required 
to drive rogue wave formation \cite{lecap12}. Because of their high amplitude or 
great steepness, rogue wave generation may be strongly affected by higher-order perturbations, such 
as those described by the Hirota equation \cite{ankiew10}, the Sasa-Satsuma equation
\cite{bandel12} and the derivative NLSE \cite{chen14}.
%

%Additional important progress has been recently obtained by extending the 
%search for rogue wave solutions to coupled-wave systems.
The study of rogue wave solutions to coupled wave systems is another hot topic, where several advances were recently reported.
Indeed, numerous physical phenomena require modeling waves
with two or more components. When compared to
scalar dynamical systems, vector systems may allow for
energy transfer between their different degrees of freedom,
which potentially yields rich and significant new families
of vector rogue-wave solutions. Rogue-wave families
have been recently found as solutions of the
vector NLSE (VNLSE)\cite{baronio12,liu2013,zhai2013,baronio14}, 
the three-wave resonant interaction equations \cite{baronio13}, 
the coupled Hirota equations \cite{chen13}, and the long-wave-short-wave 
resonance \cite{chen14R}. 

As far as rogue waves excitation is concerned, it is generally recognized that 
modulation instability (MI) is among the several mechanisms 
which may lead to rogue wave excitation. MI is a fundamental property of many 
nonlinear dispersive systems, that is associated with the growth of 
periodic perturbations on an unstable continuous-wave
background \cite{zakh09}. In the initial evolution of MI, sidebands 
within the instability spectrum experience an exponential amplification
at the expense of the pump. The subsequent wave dynamics is more complex and it
involves a cyclic energy exchange between multiple spectral modes. In fiber optics,
MI seeded from noise results in a series of high-contrast peaks of random
intensity. These localized peaks have been compared 
with similar structures that are also seen in studies of ocean rogue waves \cite{dud14}.
Nevertheless, the conditions under which MI may produce an extreme 
wave event are not fully understood. A rogue wave may be the result of MI, but conversely not every kind
of MI necessarily leads to rogue-wave generation \cite{ruderman10,sluniaev10,kharif10,baronio14}. 

%cite zakharov phys d 2009
%cite 17,18 review dudley
%2,19,20 rev dudley

%The ongoing debate has stimulated the comparison of predictions and observations between 
%hydrodynamical and optical areas in situations where analogous dynamical behaviors can be
%identified through the use of common mathematical models. 

In this work, our aim is to show that the condition for the existence of rogue wave solutions in different nonlinear wave models, 
which are commonly used both in optics and hydrodynamics, coincides with the condition of baseband MI. % Namely, rogue waves exist if and only if the MI spectrum contains the zero-frequency perturbation as a limiting case.
{\bf We define baseband MI as the condition where a cw background is unstable with respect to perturbations having infinitesimally small frequencies. Conversely, we define passband MI the situation where the perturbation experiences gain in a spectral region not including $\omega=0$ as a limiting case.}
We shall consider here the Fokas-Lenells equation (FLE) 
 \cite{chen14}, the defocusing VNLSE \cite{baronio14} and the 
long-wave-short-wave (LWSW) resonance \cite{chen14R}.
As we shall see, in the baseband-MI regime multiple 
rogue waves can be excited. Conversely, in the pass-band regime, MI only leads to the birth of
nonlinear oscillations.

{\bf We point out that, in this work, we consider as rogue wave a wave that appears from nowhere and disappears without a trace. More precisely, we take as a formal mathematical description of a rogue wave a solution that can be written in terms of rational functions, with the property of being localized in both coordinates. }
 %The fact that rogue waves have their origin in the presence of 
%baseband MI suggests the possibility to achieve a deterministic control of their excitation by superimposing to the cw pump a suitable dual-frequency input seed. In this way, MI
%develops from a coherent modulation of the input cw envelope
%rather than from random noise.

\section{Fokas-Lenells equation}
The FLE is partial differential equation that 
has been derived as a generalization of the NLSE \cite{fok95,len09}. %citare Fokas e lenells
In the context of optics, the FLE models the propagation of 
ultra-short nonlinear light pulses in monomode optical fibers \cite{len09}. 

For our studies, we write the FLE in a normalized form

\begin{equation}\label{FLNLSE}
i (1+i \kappa \partial_{\tau}) \psi_{\xi} +\frac{1}{2} \psi_{\tau \tau} +\sigma |\psi|^2 (1+i \kappa \partial_{\tau})  \psi = 0, 
\end{equation}
where $\psi (\xi,\tau)$ represents the complex envelope of the field; $\xi$, $\tau$ 
are the propagation distance and the retarded time, respectively;
each subscripted variable in Eq. (\ref{FLNLSE}) stands for partial differentiation. 
$\sigma$ ($\sigma=\pm 1$) denotes a self-focusing ($\sigma=1$) or
self-defocusing ($\sigma=-1$) nonlinearity, respectively. The real positive parameter $\kappa$ ($\kappa \geq 0$) 
represents a spatio-temporal perturbation. For $\kappa=0$, Eq. (\ref{FLNLSE}) 
reduces to the NLSE.

Soliton, multi-solitons, breathers and rogue waves solutions have been recently found 
for Eq. (\ref{FLNLSE}). Let us examine the existence condition for these rogue waves. The rogue wave solutions 
may be expressed as \cite{chen14}
\begin{equation}\label{pereFL}
\psi=\psi_0  \left[1-\frac{2i K^3 (\xi+2 \kappa \tau)+ \sigma K/ a^2)}{D+ i \kappa K \gamma}\right] 
\end{equation}
where $\psi_0=a e^{i (\omega \tau - \beta \xi)}$
represents the background solution of Eq. (\ref{FLNLSE}),
$a$ is the real amplitude parameter ($ a>0$),
$\omega$ the frequency; moreover $\beta=\omega^2/2K-\sigma a^2$, $K=1-\omega \kappa$, 
$\gamma=K^2 \tau + (K^2-1)\xi/(2 \kappa)$, 
$D=(\sigma \gamma+ a^2 \kappa K \xi)^2+ a^2 \alpha^2 \xi^2+ \sigma K / (4a^2)$, 
$\alpha=\pm \sqrt{\sigma K- a^2 \kappa^2 K^2}$.

\begin{figure}[h!]
\centerline{\includegraphics[width=7cm]{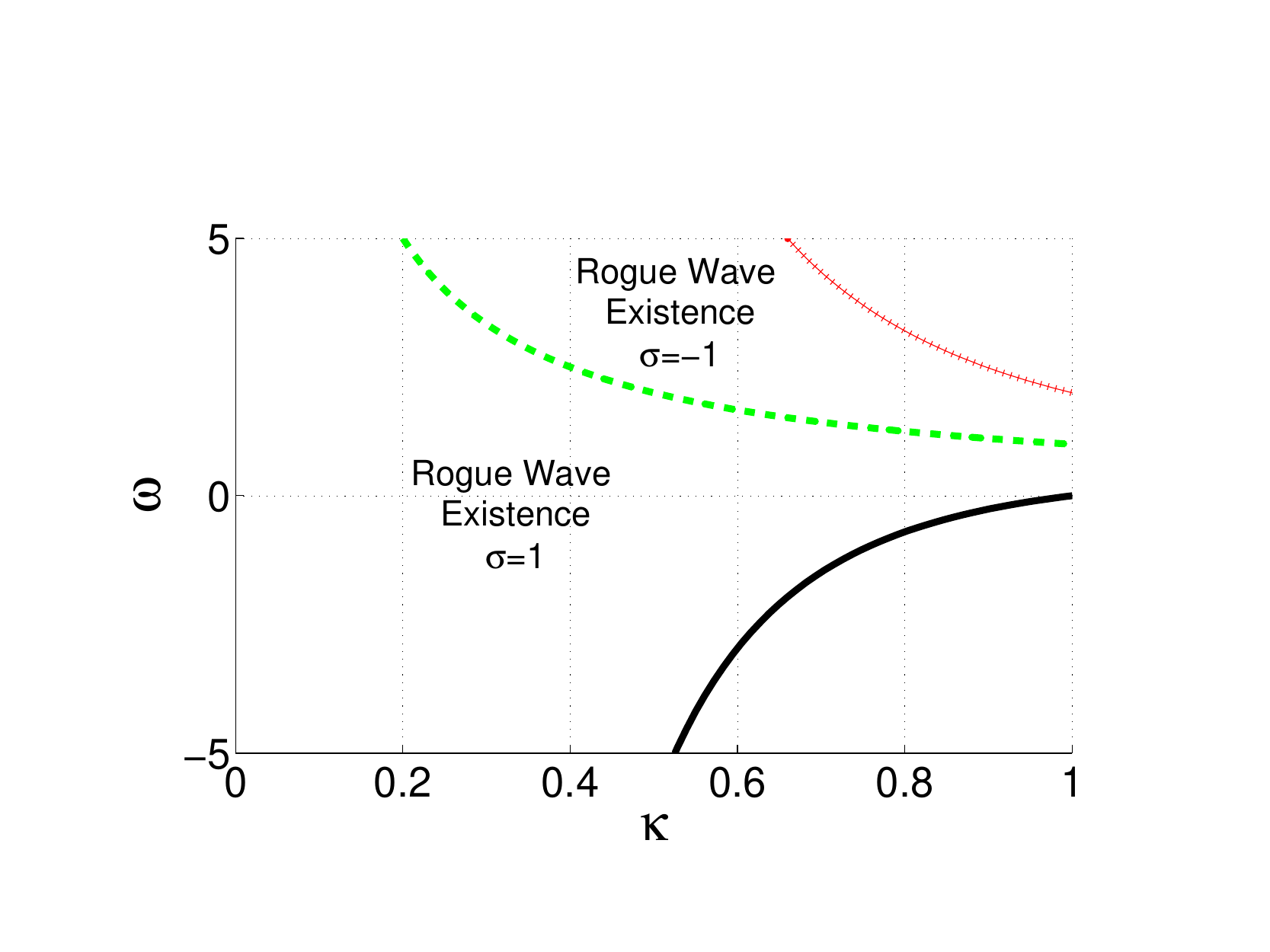}}
     \caption{Existence domains of rogue waves in the plane $(\kappa,\omega)$,
		with $a=1$, in the focusing regime ($\sigma=1$)
		and defocusing regime ($\sigma=-1$). The red dotted line denotes
		$\omega=1/\kappa+1/\kappa^3$; green dashed line $\omega=1/\kappa$;
		black solid line $\omega=1/\kappa-1/\kappa^3$.} \label{fig_0}
\end{figure}

\begin{figure}[h]
\centerline{\includegraphics[width=7cm]{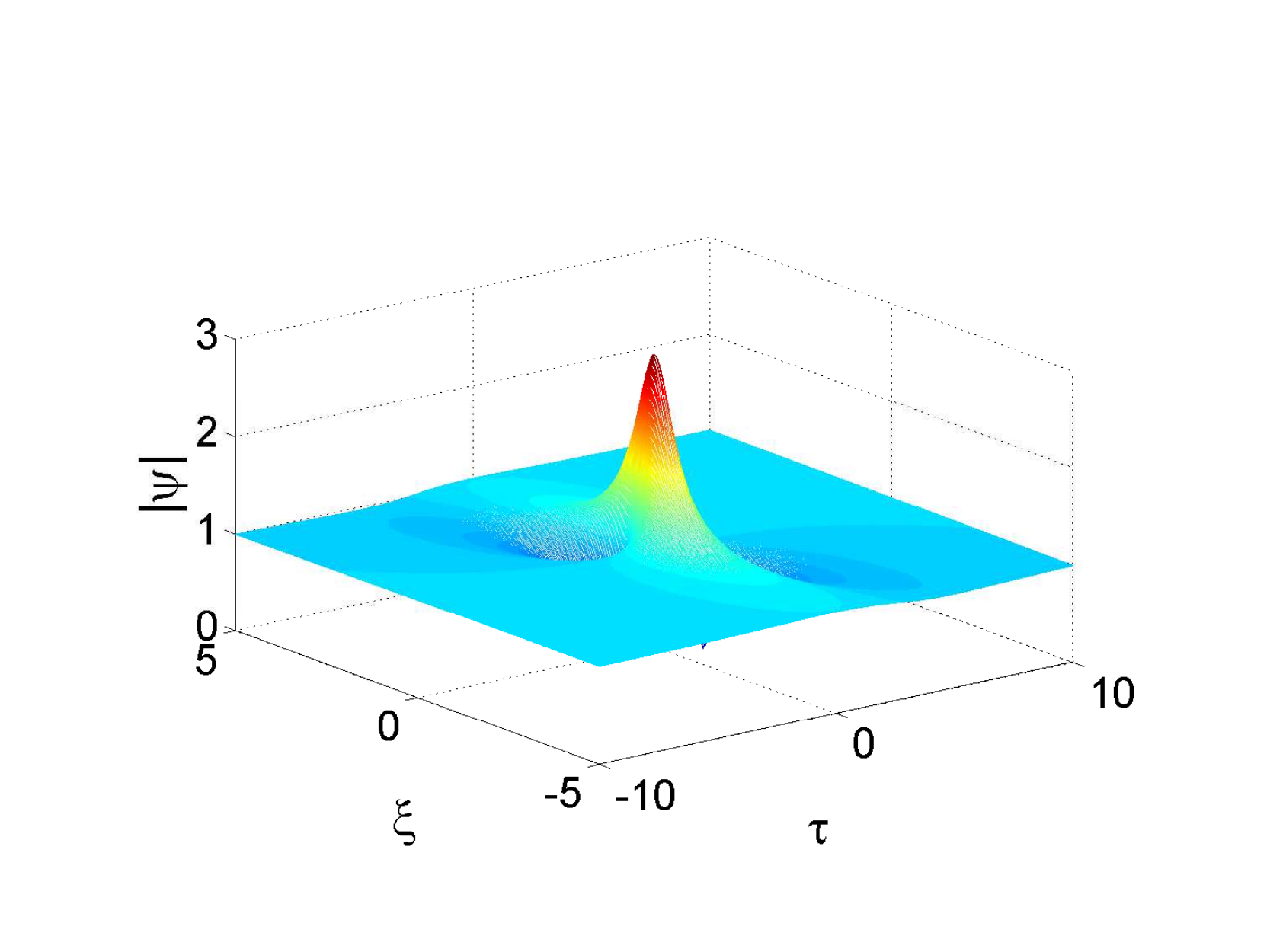}}
\centerline{\includegraphics[width=7cm]{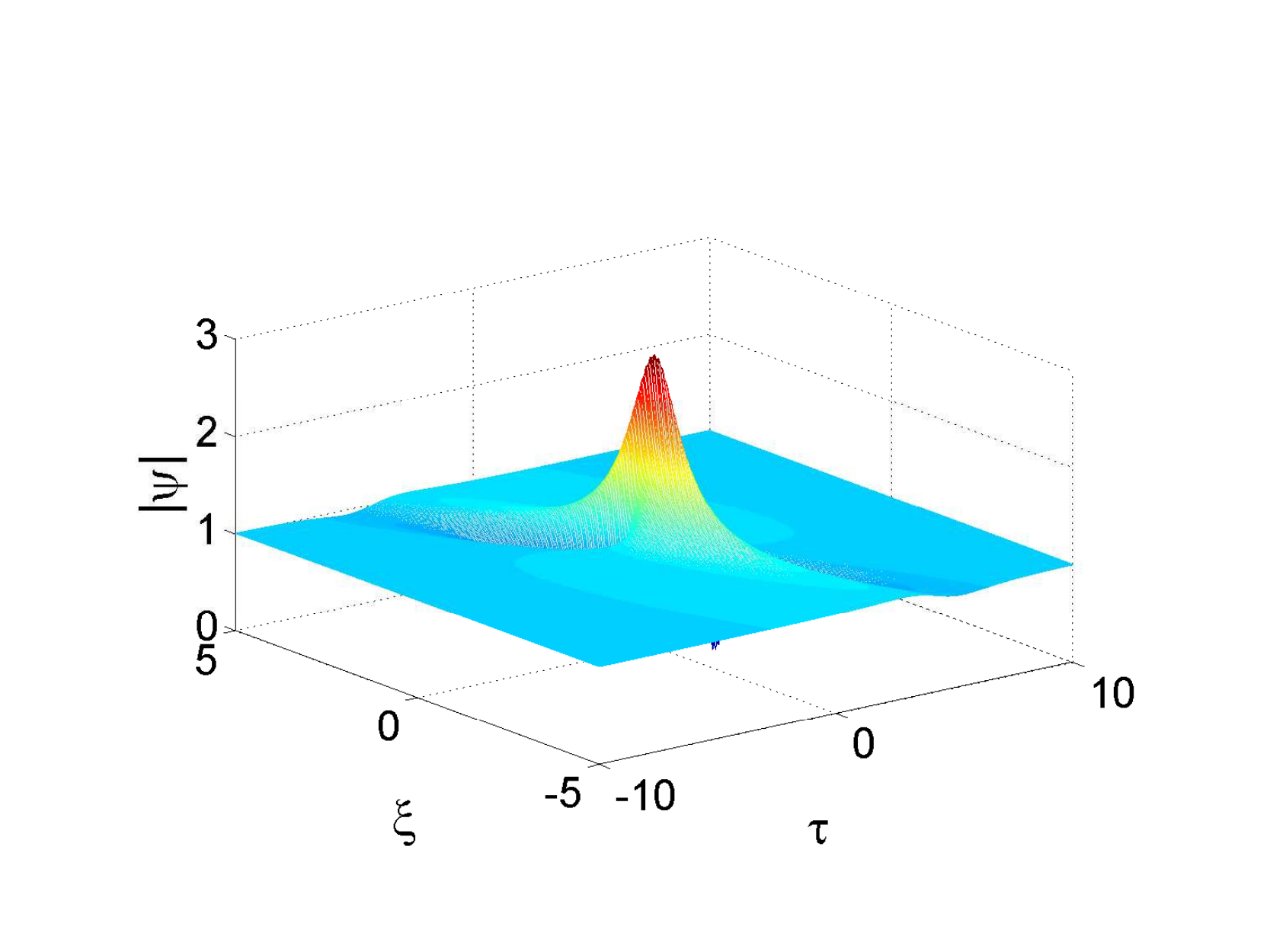}}
     \caption{Typical rogue soliton states. Top, focusing regime $\sigma=1$, $\kappa=0.5$ 
		and $a=1, \omega=0$. Bottom, defocusing regime $\sigma=-1$, $\kappa=0.5$ 
		and $a=1, \omega=4$. } \label{fig_1}
\end{figure}

The rogue wave solutions (\ref{pereFL}) depend on the real parameters $a$ and $\omega$,
for fixed $\sigma$ and $\kappa$. In the focusing regime ($\sigma=1$), rational rogue waves exist for
$\omega$ in the range $[1/\kappa-1/(a^2 \kappa^3), 1/\kappa]$. Whereas in the defocusing regime
rogue waves exist for $\omega$ in the range $[1/\kappa, 1/\kappa+1/(a^2 \kappa^3)]$.
Figure \ref{fig_0} shows the domains of rogue wave existence in the plane
$(\omega,\kappa)$, for either the focusing or the defocusing regimes. 
Surprisingly,  exponential soliton
states exist in the complementary region of the $(\omega, \kappa)$ plane (see Ref. \cite{chen14} for details on the properties of these nonlinear waves).
Figure \ref{fig_1} illustrates a typical example of rogue wave solution (\ref{pereFL}). 


Let us turn our attention now to the linear stability analysis of the background
solution of Eq.(\ref{FLNLSE}). A perturbed nonlinear background can be written 
as $\psi_p=[a+p] e^{i (\omega \tau - \beta \xi)}$,
where $p(\xi,\tau)$ is a small complex perturbation that
satisfies a linear differential evolution equation. Whenever $p$ is $\tau$-periodic with frequency $\Omega$,
i.e., $p(\xi,\tau)=\eta_s(\xi) e^{i\Omega \tau} + \eta_a(\xi) e^{-i\Omega \tau}$,
such equation reduces to a set of $2 \times 2$ linear ordinary differential equations 
$\eta'=iM \eta$, with $\eta=[\eta_{s},\eta^*_{a}]^T$
(here a prime stands for differentiation with respect to $\tau$). For any given real frequency $\Omega$,
the generic perturbation $\eta(\xi)$ is a linear combination of exponentials $e^{i w_j \xi}$
where $w_j, (j=1,2)$ are the two eigenvalues of the matrix $M=\{M_{ij}\}$, whose elements read as:

\begin{align*}
M_{11}&=\frac{-\frac{1}{2} \Omega^2+\sigma a^2K-\Omega (\omega+ \beta \kappa+\sigma a^2 \kappa)}{ (K-\kappa \Omega)},\\
M_{12}&=\frac{\sigma a^2 K}{(K-\kappa \Omega)},\\
M_{21}&=-\frac{\sigma a^2 K}{ (K+ \kappa \Omega)},\\
M_{22}&=\frac{\frac{1}{2} \Omega^2-\sigma a^2K-\Omega (\omega+ \beta \kappa+\sigma a^2 \kappa)}{ (K+\kappa \Omega)}.
\end{align*}

Since the entries of the matrix
M are all real,  
%$M_{11}=[-\frac{1}{2} \Omega^2+\sigma a^2K-\Omega (\omega+ \beta \kappa+\sigma a^2 \kappa)]/ (K-\kappa \Omega)$,
%$M_{12}=\sigma a^2 K / (K-\kappa \Omega)$, $M_{21}=-\sigma a^2 K / (K+ \kappa \Omega)$,
%$M_{22}=[\frac{1}{2} \Omega^2-\sigma a^2K-\Omega (\omega+ \beta \kappa+\sigma a^2 \kappa)]/ (K+\kappa \Omega)$,
the eigenvalues $w_j$ are either real or they appear as complex conjugate pairs. The eigenvalues of the matrix $M$ are the roots of its
characteristic polynomial, 
\begin{align}
\label{polyc}B(w)&= B_2 w^2 +B_1 w + B_0,\\
\nonumber B_2&=K^2-\kappa^2\Omega^2,\\
\nonumber B_1&=-4 \Omega (2 \beta \kappa K +\kappa \Omega^2 +2 K \omega),\\
\nonumber B_0&=-\Omega^2+4 [\beta^2 \kappa^2+a^4\kappa^2+\omega^2+\\
\nonumber   &2 \beta \kappa (a^2\kappa \sigma +\omega)
+A^2 \sigma (K+2 \kappa \omega).
\end{align}

 Mi occurs whenever M has an eigenvalue $w$ with a negative imaginary part. Indeed, if the explosive rate is 
 $G(\Omega)=-\textrm{Im}\{w\} >0$, perturbations grow exponentially like $\exp(G \xi)$ at the expense of the pump wave. 

\begin{figure}[h]
\begin{center}
\includegraphics[width=7.6cm]{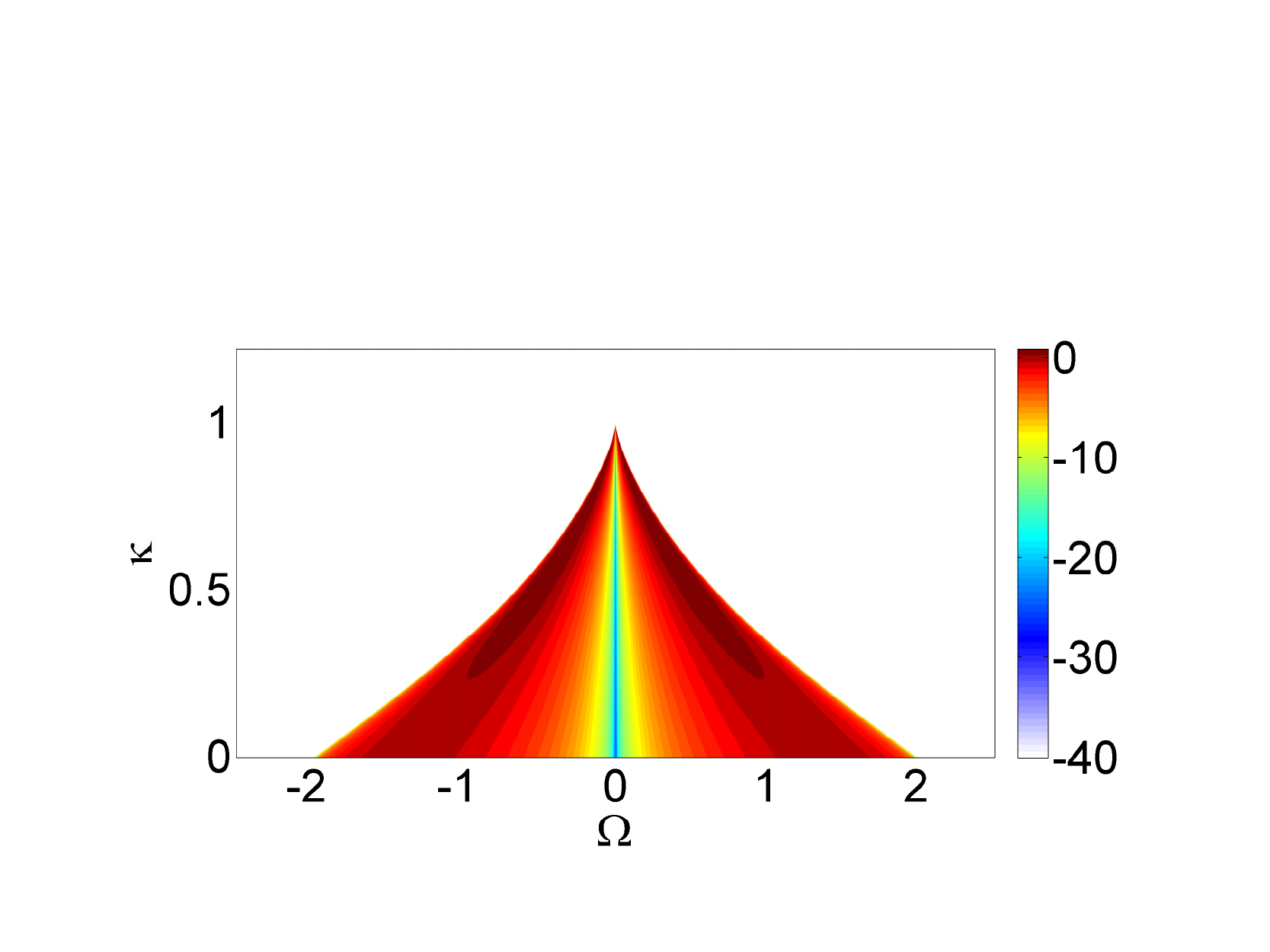}
\includegraphics[width=7.6cm]{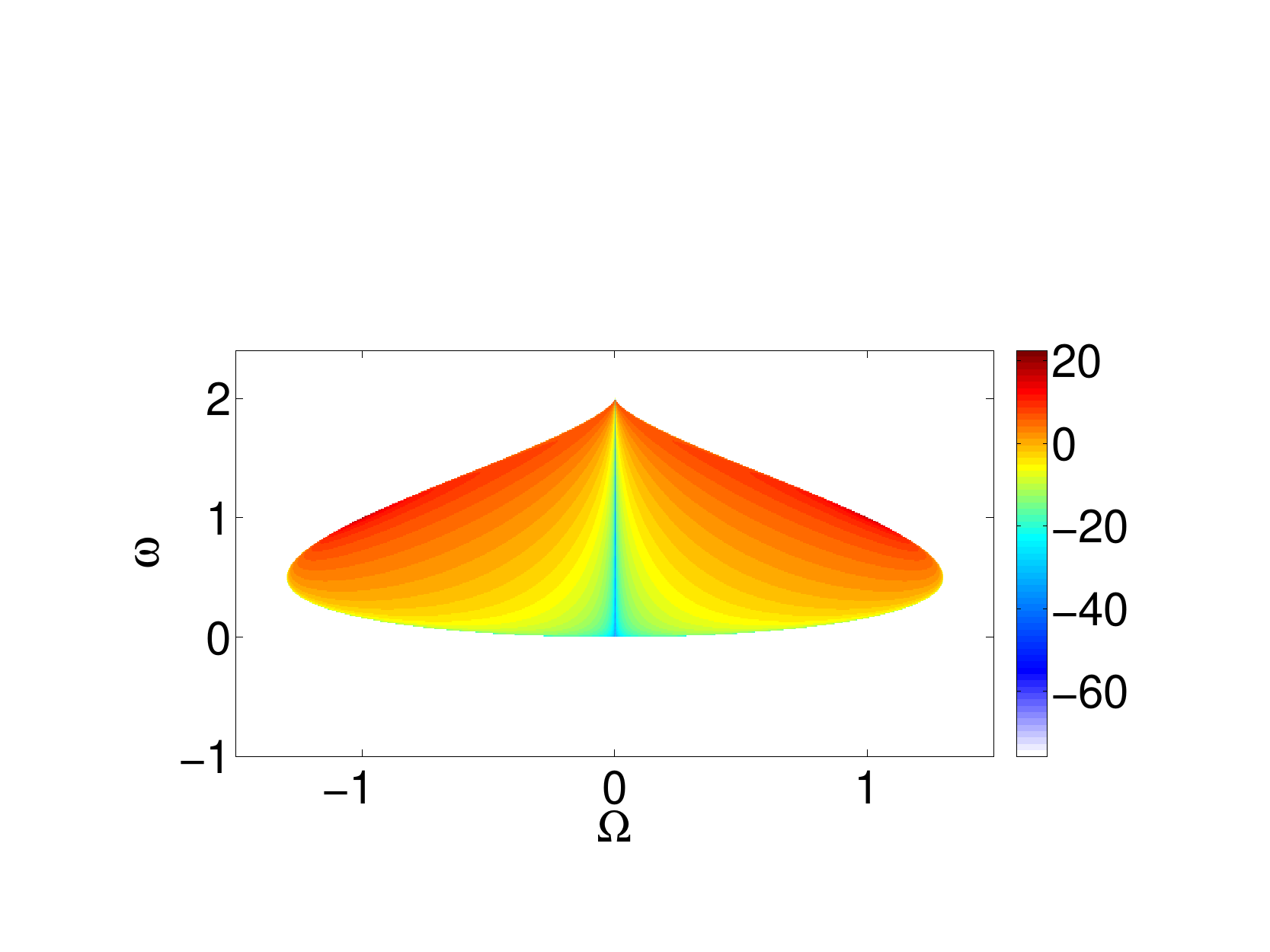}
    \end{center}
     \caption{Maps of logaritmic MI gain ($10log_{10} G$) in the focusing ($\sigma=1$) FLE (\ref{FLNLSE}). 
		Top, MI on the $(\Omega,\kappa)$ plane, calculated for the case $a=1, \omega=1$.
		Bottom, MI on the $(\Omega,\omega)$ plane, calculated for the case $a=1, \kappa=0.5$. 
			} \label{fig_2a}
\end{figure}

\begin{figure}[h]
\begin{center}
\includegraphics[width=7.6cm]{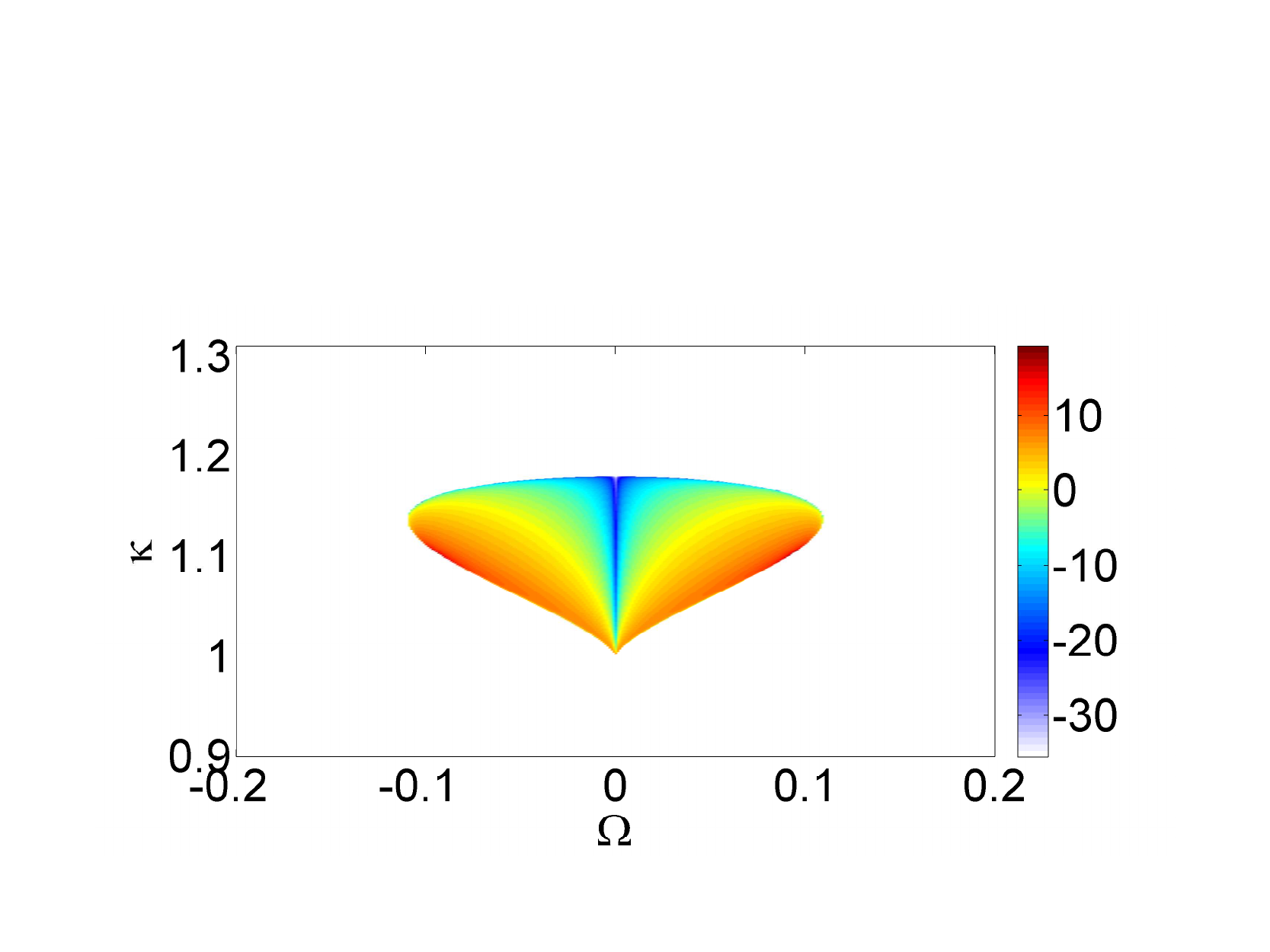}
    \end{center}
     \caption{Maps of of logaritmic MI gain ($10log_{10} G$)  in the defocusing ($\sigma=-1$) FLE (\ref{FLNLSE}). 
		MI on the $(\Omega,\kappa)$ plane, calculated for the case $a=2, \omega=1$. } \label{fig_2b}
\end{figure}

MI is well depicted by displaying the gain $G(\Omega)$ as function of $a, \omega, \sigma, \kappa$ and $\Omega$.
The resulting MI gain spectrum is illustrated in Fig. \ref{fig_2a} and Fig. \ref{fig_2b}. 
These figures show the MI gain in the focusing ans defocusing regime, respectively. In both cases, baseband MI is only present in a certain subset of the $\omega, \kappa$ parameters.
 Since the gain band (where $G(\Omega)\neq 0$) can be written as  $0 \leq \Omega_1 < \Omega < \Omega_2$ (and its symmetric counterpart with respect to $\Omega=0$), baseband MI
is obtained if $\Omega_1=0$, whereas passband MI occurs for $\Omega_1>0$. 

We proceed next by focusing our attention on the MI gain spectrum, by evaluating the sign of the discriminant $\Delta$ of the characteristic polymomial (\ref{polyc}): this leads to
\begin{equation}\label{disc}
sign\{\Delta\}=sign\{\Omega^2-4a^2\sigma K^3\left(1-a^2\kappa^2\sigma K\right)\}.
\end{equation}

 %\textit{sign}$\left\{\Delta\right\}=(a^2 \kappa^2 K^2-\sigma K)$.  
If the discriminant $\Delta$ is positive, the characteristic polynomial has
two real roots and there is no MI. On the other hand if the discriminant $\Delta$ is negative, the characteristic polynomial $B$ has two complex conjugate roots, and Eq. (\ref{FLNLSE}) exhibits baseband MI. It is clear from Eq. (\ref{disc}) that for FLE if there is MI, it is of baseband type only: either the system is modulationally unstable for $\Omega\rightarrow 0$, either there is no MI at all.
The interesting finding is that the sign constraint on the discriminant, 
which determines the presence of baseband MI, 
leads to the condition that $\omega$ should be in the range $[1/\kappa-1/(a^2 \kappa^3), 1/\kappa]$ 
in the focusing regime ($\sigma=1$), and
in the range  $[1/\kappa, 1/\kappa+1/(a^2 \kappa^3)]$ in the defocusing regime ($\sigma=-1$).
These conditions exactly coincide with the constraints that are required for the existence of the rogue wave solution (\ref{pereFL}).

These results are important since they show that, for both the focusing and the defocusing regime, rogue wave solutions of Eq.  (\ref{FLNLSE}) only exist in the subset of the parameters 
space where also baseband MI is present.

\begin{figure}[h]
\begin{center}
\includegraphics[width=7.cm]{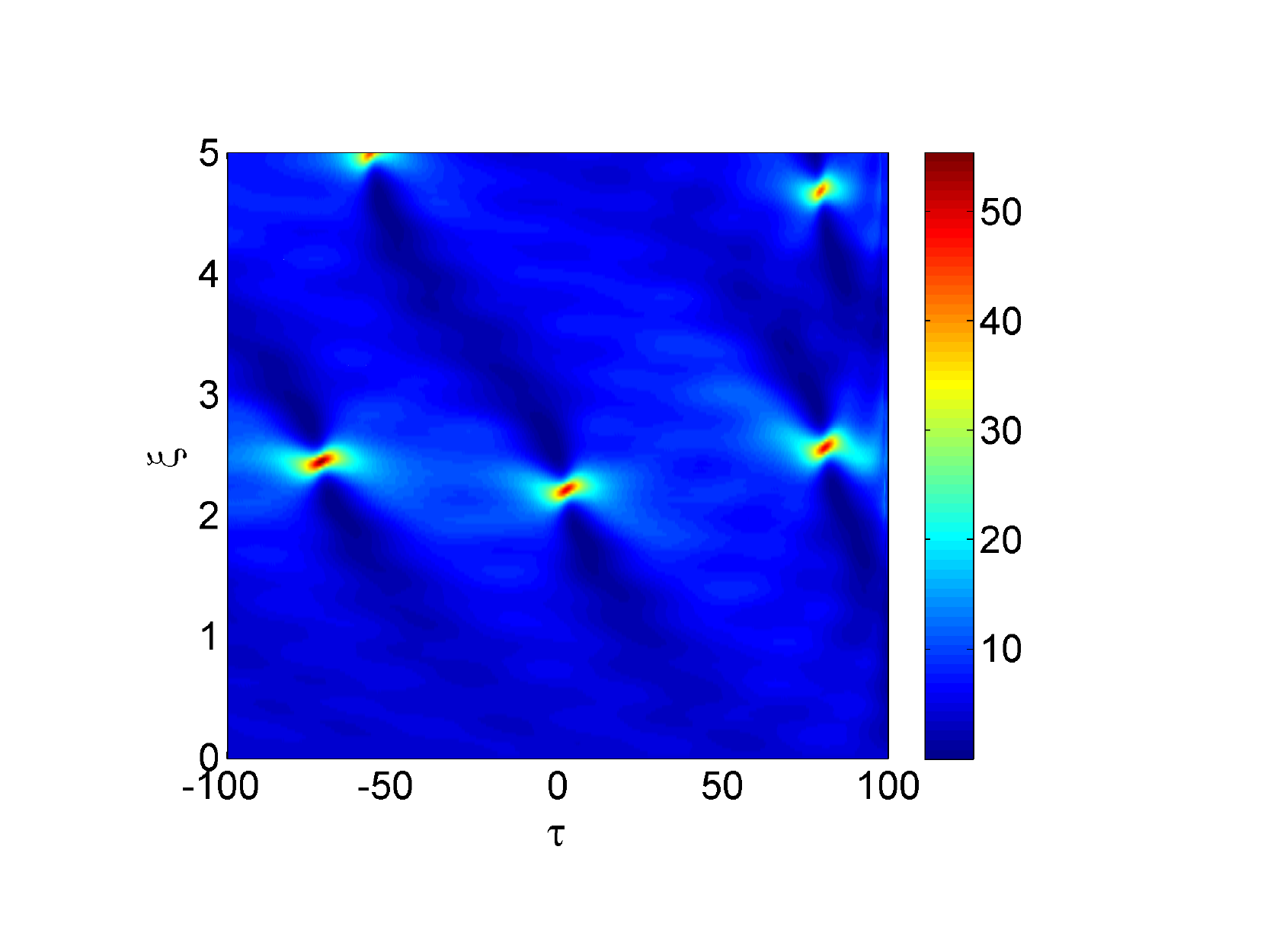}
    \end{center}
     \caption{Color plot of $|\psi(\xi,\tau)|^2$ from the numerical solution of the focusing FLE (\ref{FLNLSE}) in the baseband MI
regime. The initial condition is a plane wave perturbed by a random noise seed, with
parameters: $ a= 2, \omega=1$, $\kappa=1.15, \sigma=-1$.} \label{f3}
\end{figure}

%The fact that rogue waves have their origin in baseband MI suggests the possibility
%to control their excitation using a dual frequecy input field such that the 
%instability develops from a coherent modulation of the input envelope
%rather than from noise.
We checked the results of our analysis by extensive numerical solutions of Eq. (\ref{FLNLSE}). 
These simulations indeed confirm that, in the baseband MI regime, multiple rogue waves can generated from
an input plane wave background with a superimposed random noise seed (see Fig. \ref{f3}).

%\begin{figure}[h]
%\begin{center}
%\includegraphics[width=7.cm]{f3.eps}
    %\end{center}
     %\caption{Color plot of $|\psi(\xi,\tau)|$ from numerical solution of focusing (\ref{FLNLSE}), in the baseband MI
%regime. Initial condition is a plane wave perturbed by a small $\tau$-periodic coherent modulation. 
%Parameters: $ a= 1, \omega=0$, $\kappa=0.5, \sigma=1$. Coherent modulation is inside the MI band.} \label{f3}
%\end{figure}
%%

\section{Defocusing VNLSE}

The defocusing VNLSE constitutes another model that has been thoroughly exploited for the description of fundamental physical phenomena in several different disciplines.
In oceanography, for instance, it may describe the interaction of crossing currents \cite{miguel_crossing}. In the context of 
nonlinear optics, it has been derived for the description of pulse propagation in randomly birefringent fibers \cite{meniuk}, or coupled beam propagation in photorefractive media \cite{segev}.

%
%Intor di applicazione defocusing
%
%
For our studies, we write the defocusing VNLSE in the following dimensionless form

\begin{equation}\label{VNLS}
\left \{ \begin{array} {lll}
i\psi^{(1)}_\xi+ \psi^{(1)}_{\tau \tau} -2 ( |\psi^{(1)}|^2 +  |\psi^{(2)}|^2 ) \psi^{(1)} & = & 0 \\
i\psi^{(2)}_\xi+ \psi^{(2)}_{\tau \tau} -2 ( |\psi^{(1)}|^2  + |\psi^{(2)}|^2 ) \psi^{(2)} & = & 0, 
\end{array} \right .
\end{equation}
where $\psi^{(1)}(\xi,\tau),\,\psi^{(2)}(\xi,\tau)$ represent complex wave envelopes; $\xi,\tau$ are the propagation
distance and the retarded time, respectively; each subscripted variable in Eqs. (\ref{VNLS}) stands for partial differentiation. 
Note that Eqs. (\ref{VNLS}) refer to the defocusing (or normal dispersion) regime.
Unlike the case of the scalar NLSE, rational rogue solutions
of the defocusing VNLSE do exist, as it was recently demonstrated \cite{baronio14}. 
These rogue wave solutions can be expressed as:

%\begin{subequations}\label{pere}
\begin{equation}\label{pere}
\psi^{(j)}=
  \psi_0^{(j)} \big[\frac{p^2 \tau^2+p^4 \xi^2+p \tau(\alpha_j+\beta \theta_j)-i \alpha_j p^2 \xi+ \beta \theta_j}{p^2\tau^2+p^4\xi^2+ \beta (p\tau+1)}  \big] 
\end{equation}
with $j=1,2$. $\psi_0^{(j)} =a_j e^{i(\omega_j \tau- \beta_j \xi)}$, represent the background solution of 
Eqs. (\ref{VNLS}), $a_j$ are the real amplitude parameters ($a_j>0$), $\omega_j$ are the frequencies, and
$\beta_j=\omega_j^2+ 2 (a_1^2+a_2^2)$. 
%\end{gather*}

% 
%
Moreover, $\alpha_j=4p^2/(p^2+4\omega_j^2), \theta_j=(2\omega_j+ip)/(2\omega_j-ip); 
\beta=p^3/\chi(p^2+4\omega_1 \omega_2), p=2 \textrm{{Im}}(\lambda+k), 
\omega_1+\omega_2=2 \textrm{{Re}}(\lambda+k),  \omega_1-\omega_2=2\omega, \chi=\textrm{{Im}} k.$ 
The evaluation of the complex value of $\lambda$ and $k$ should be performed as follows.
The parameter $\lambda$ is the double solution of the polynomial 
$A(\lambda)=\lambda^3+ A_2 \lambda^2+ A_1 \lambda +A_0=0$, 
with
$A_0=-k^3 + k(\omega^2 + a_1^2 + a_2^2) + \omega (a_2^2 -a_1^2)$,
$A_1 = -k^2 - \omega^2 + a_1^2 + a_2^2$, $A_2 = k$.
Moreover, the constraint on the double roots of $A(\lambda)$ is satisfied whenever 
the discriminant of $A(\lambda)$ is zero, which results in
the fourth order polinomial condition $D(k)=k^4+D_3 k^3+ D_2 k^2+ D_1 k +D_0=0$,
with %
$D_0 = (\omega^2 - a_1^2 - a_2^2)^3/(2^4 \omega^2) - (3/4)^3 (a_2^2 - a_1^2)^2$,
$D_1 = -9(a_2^2 - a_1^2)(2 \omega^2 + a_1^2 + a_2^2)/(2^4 \omega)$,
$D_2 = -[8q^4 - (a_1^2 +a_2^2)^2 +20 \omega^2(a_1^2 +a_2^2)]/(2^4 \omega^2)$,
$D_3 = (a_2^2 - a_1^2)/(2\omega)$.
Thus, $\lambda$ is the double solution of the third order polynomial $A(\lambda)$,
and $k$ is any strictly complex solution of the fourth order polynomial $D(k)$
(see Ref. \cite{baronio14} for details on nonlinear waves calculations and characteristics).

\begin{figure}[h]
\begin{center}
\includegraphics[width=7cm]{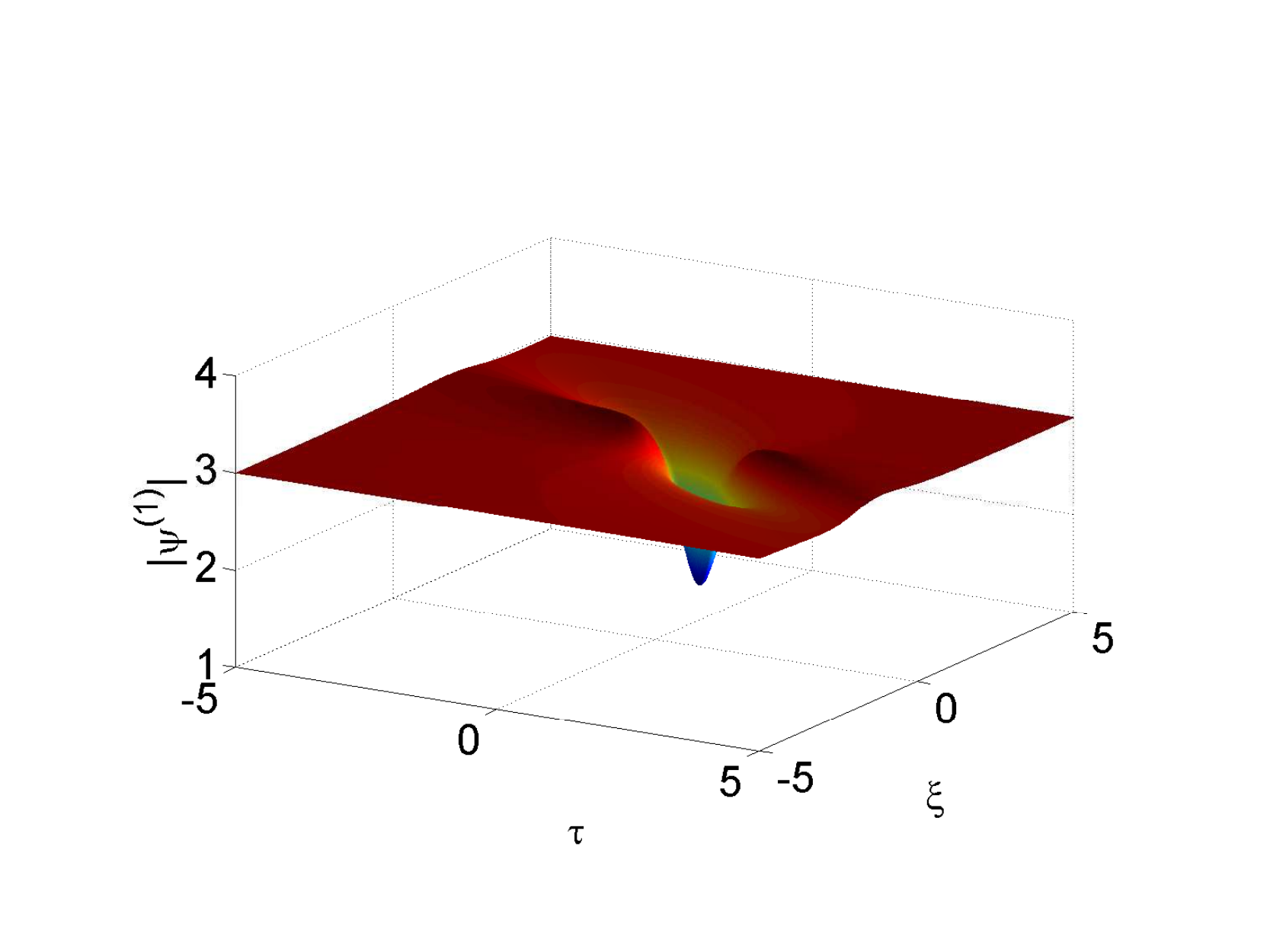}
\includegraphics[width=7cm]{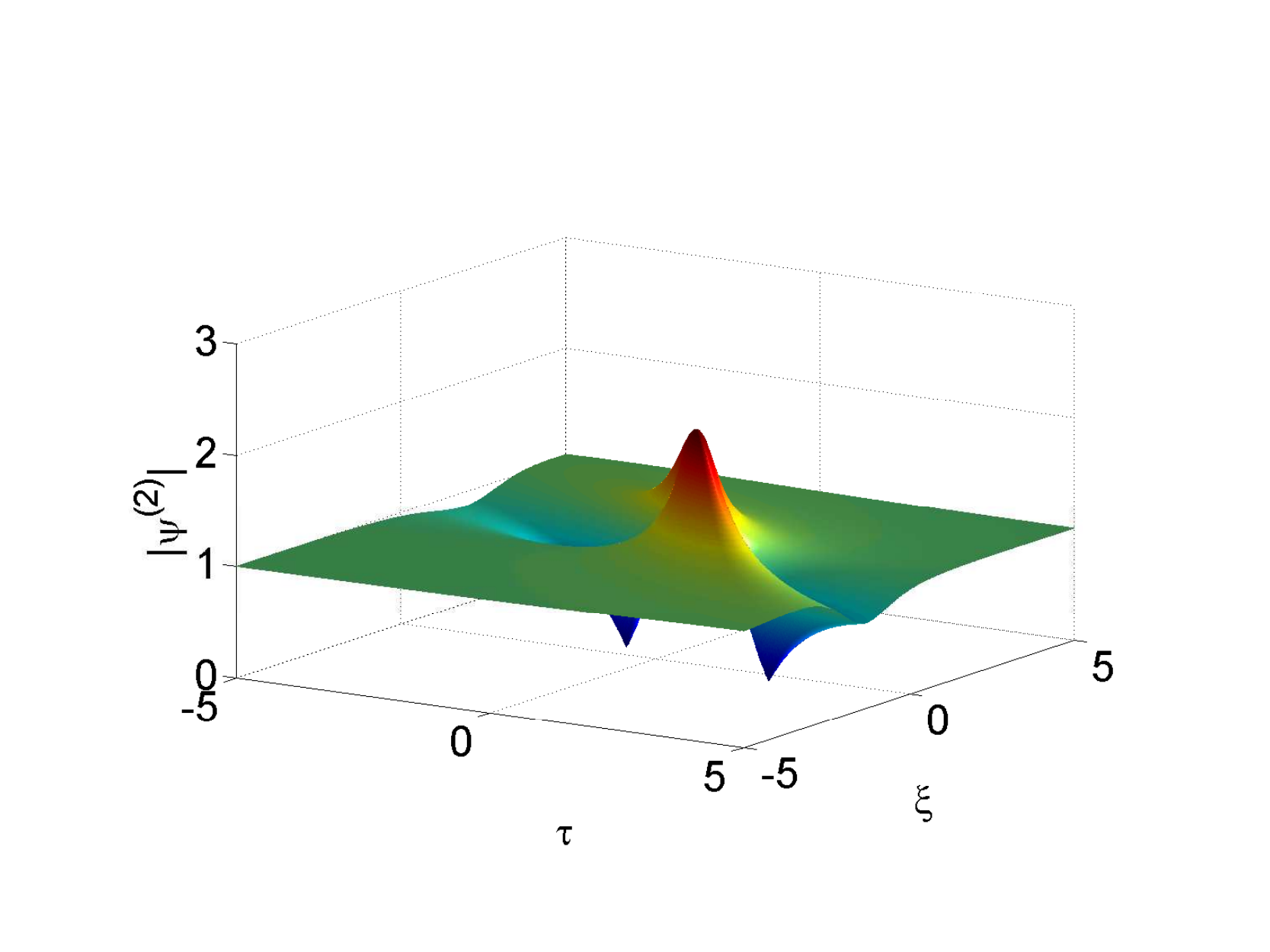}
    \end{center}
     \caption{Rogue wave envelope distributions $|\psi^{(1)}(\tau,\xi)|$
     and $|\psi^{(2)}(\tau,\xi)|$ of expression (\ref{pere}). Here, $a_1=3, a_2=1, \omega=1$.
		$k=2.36954 + 1.1972i$ and $\lambda=-1.69162 - 1.79721i$.
    } \label{figv_1}
\end{figure}

The rogue waves (\ref{pere}) depend on the real parameters $a_1, a_2$ and $\omega$ {which} originate from the backgrounds:
$a_1, a_2$ represent the amplitudes, and $2\omega$ the ``frequency'' difference of the waves. 
Figure \ref{figv_1} shows a typical dark-bright solution (\ref{pere}).

\begin{figure}[h]
\begin{center}
\includegraphics[width=4.2cm]{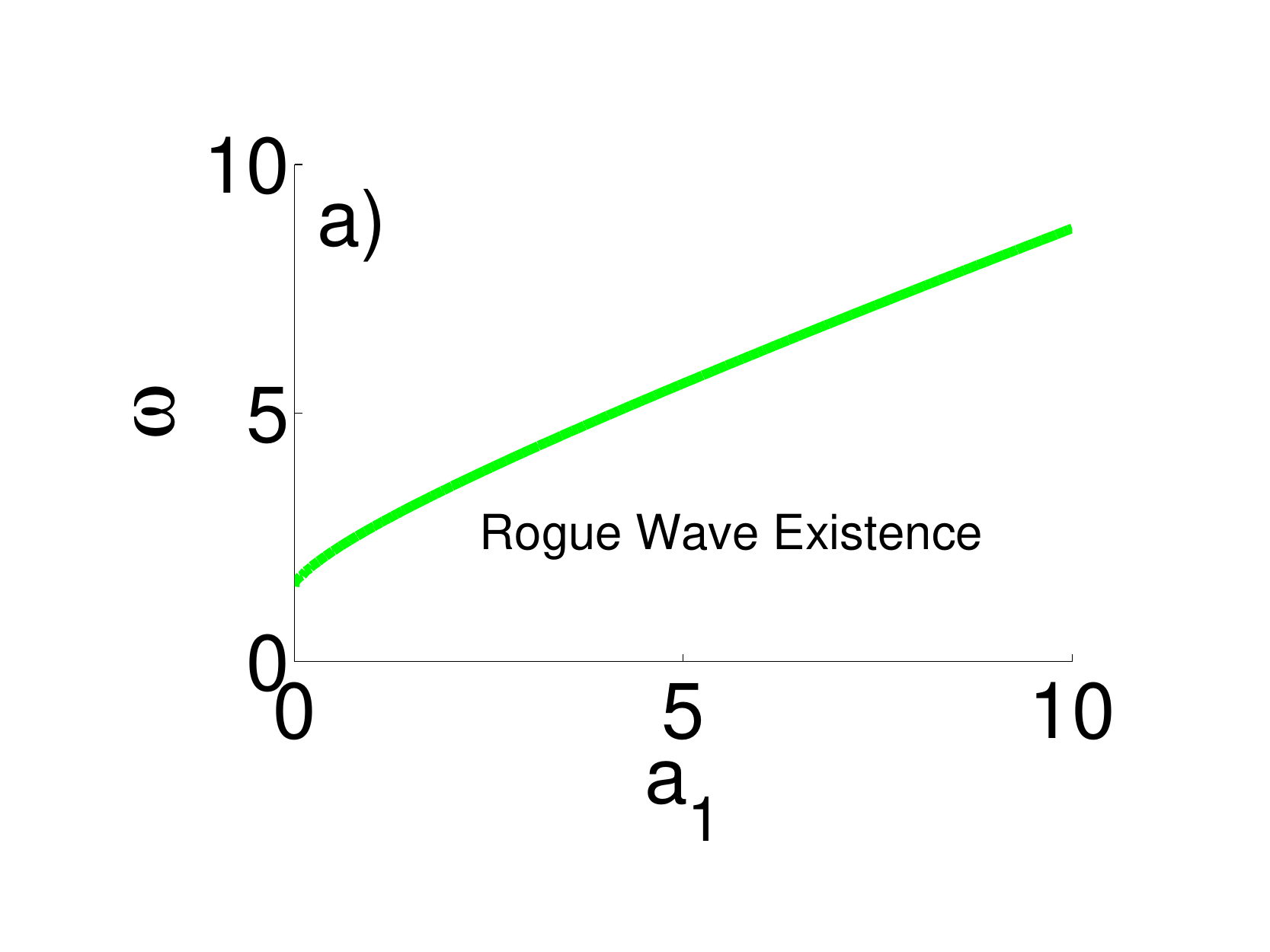}
\includegraphics[width=4.2cm]{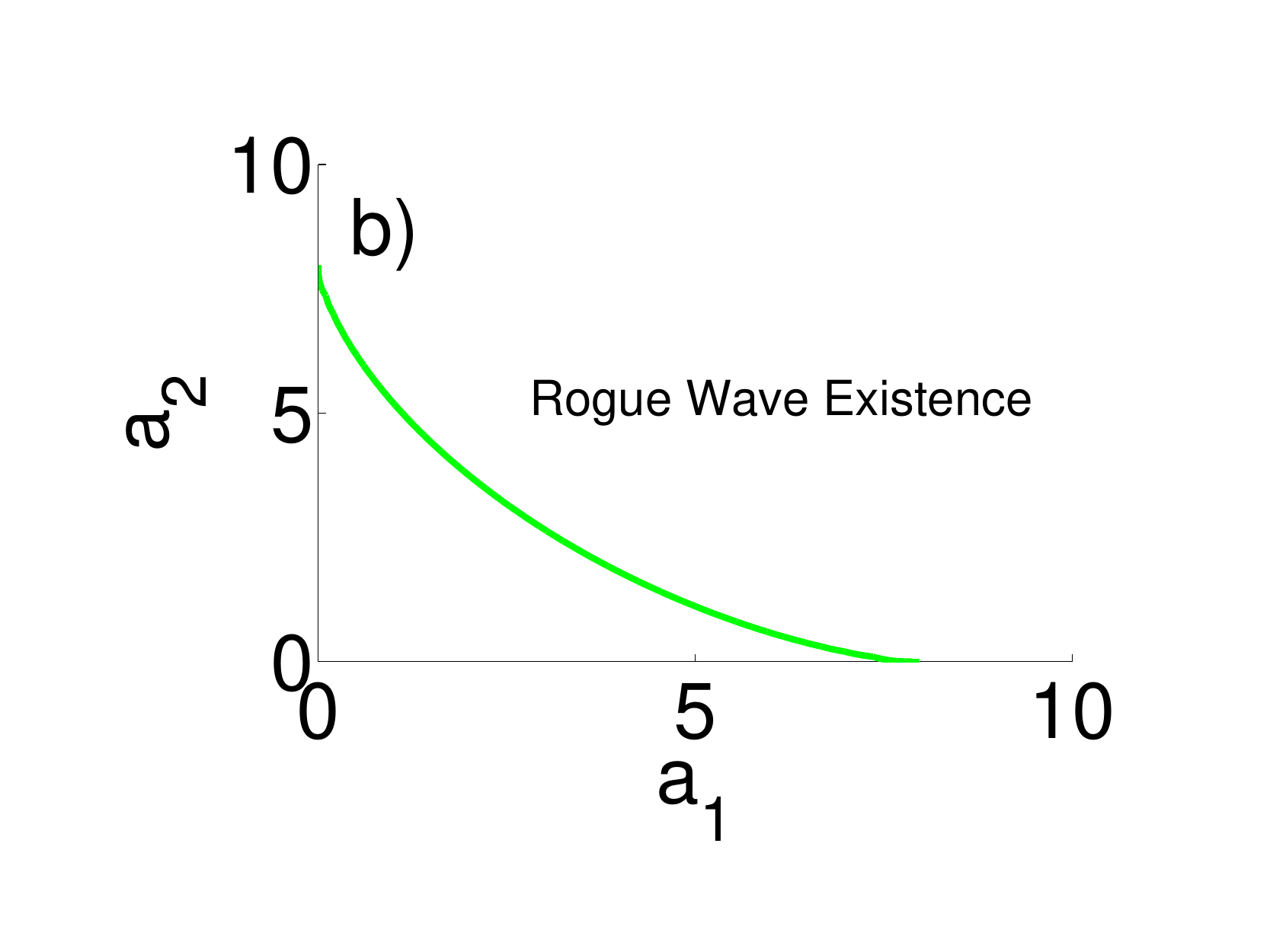}
    \end{center}
     \caption{Rogue wave existence condition. a) $(\omega,a_1)$ plane,
		with $a_2=3$. b) $(a_2,a_1)$ plane, with $\omega=4$.
    } \label{fignc}
\end{figure}

In the defocusing regime, it has been demonstrated \cite{baronio14} that  rogue waves exist in the subset of parameters
$a_1, a_2, \omega$ where
\begin{equation} \label{eqkprima}
(a_1^2+a_2^2)^3 -12 (a_1^4 -7 a_1^2 a_2^2 +a_2^4) \omega^2 + 48 (a_1^2 + a_2^2) \omega^4 - 64 \omega^6 >0.
\end{equation}
Figure \ref{fignc} illustrates two characteristic examples of the existence condition for rogue waves. In particular,
Fig.\ref{fignc} shows that, for a fixed $\omega$, the background amplitudes should be sufficiently large
in order to allow for rogue wave formation. 

{Let us  turn our attention now to the linear stability analysis of the background solution of Eqs.(\ref{VNLS}).
A perturbed nonlinear background may be written as 
$\psi_p^{(j)}=  [a_j+p_j] e^{i \omega_j \tau-i \beta_j \xi}$, where $p_j(\xi,\tau)$ are small complex perturbations that obey a linear partial differential equation.  Whenever $p_j(\xi,\tau)$ are $\tau-$periodic with frequency $\Omega$, i.e.,
$p_j(\xi,\tau)=\eta_{j,s}(\xi)e^{i \Omega \tau}+\eta_{j,a}(\xi)e^{-i \Omega \tau}$, their equations reduce to the $4\times4$ linear ordinary differential equation $\eta'=iM\eta$, with $\eta=[\eta_{1,s},\eta^*_{1,a},\eta_{2,s},\eta^*_{2,a}]^T$ . For any given real frequency $\Omega$, the generic perturbation $\eta(\xi)$ may be expressed by a linear combination of exponentials $\exp(i w_j \xi)$ where $w_j,\;j=1,\cdots,4, $ are the four eigenvalues of the matrix $M=\{M_{ij}\}$.

\begin{align*}
M_{11}&=-\Omega^2-2 \Omega \omega_1 -2a_1^2,\\ 
M_{22}&=\Omega^2-2\Omega\omega_1+2a_1^2,\\ 
M_{33}&=-\Omega^2-2\Omega\omega_2-2a_2^2,\\
M_{44}&=\Omega^2-2\Omega\omega_2+2a_2^2,\\
M_{12}&=-M_{21}=-2a_1^2,\\
M_{13}&=M_{14}=M_{31}=M_{32}=-2a_1 a_2,\\
M_{41}&=M_{23}=M_{24}=M_{42}=  2a_1 a_2,\\
M_{43}&=-M_{34}=2a_2^2.
\end{align*}

 Since the entries  of the matrix $M$ are all real, 
%$M_{11}=-\Omega^2-2 \Omega \omega_1 -2a_1^2 $, 
%$M_{22}=\Omega^2-2\Omega\omega_1+2a_1^2$, $M_{33}=-\Omega^2-2\Omega\omega_2-2a_2^2$, $M_{44}=\Omega^2-2\Omega\omega_2+2a_2^2$,
%$M_{12}=-M_{21}=-2a_1^2$,
%$M_{13}=M_{14}=M_{31}=M_{32}=-M_{41}=-M_{23}=-M_{24}
%=-M_{42}= -2a_1 a_2 $, $M_{43}=-M_{34}=2a_2^2$,
the eigenvalues $w_j$ are either real or they appear as complex conjugate pairs. These eigenvalues are the roots of the characteristic polynomial $B(w)$ of the matrix $M$: 
\begin{align*}
B(w)&=w^4+B_3 w^3+ B_2 w^2+ B_1 w +B_0,\\
B_0 &= (\Omega^2 - 4 \omega^2  ) [4 (a_1^2 +  a_2^2 - \omega^2) + \Omega^2] \Omega^4,\\ 
B_1 &=  16 \omega (a_1^2 - a_2^2)  \Omega^3,\\ 
B_2 &= - 2  [2 (a_1^2 + a_2^2 + 2 \omega^2) + \Omega^2] \Omega^2,\\
B_3 &= 0.
\end{align*} 
MI occurs whenever $M$ has an eigenvalue $w$ with a negative imaginary part, $\textrm{Im}\{w\} < 0$. 
Indeed, if the explosive rate is $G(\Omega) = -\textrm{Im}\{w\} >0$, initial perturbations  
grow exponentially as $\exp(G \xi)$ at the expense of the pump waves. 
%
%MI is well depicted by displaying the  gain $G(\Omega)$ as a function  of $a_1, a_2$, $\omega$, $\Omega$. 
Typical shapes of the MI gain $G(\Omega)$ are shown in  Fig. \ref{fig_2}. 
\begin{figure}[h]
\begin{center}
\includegraphics[width=8cm,height=4cm]{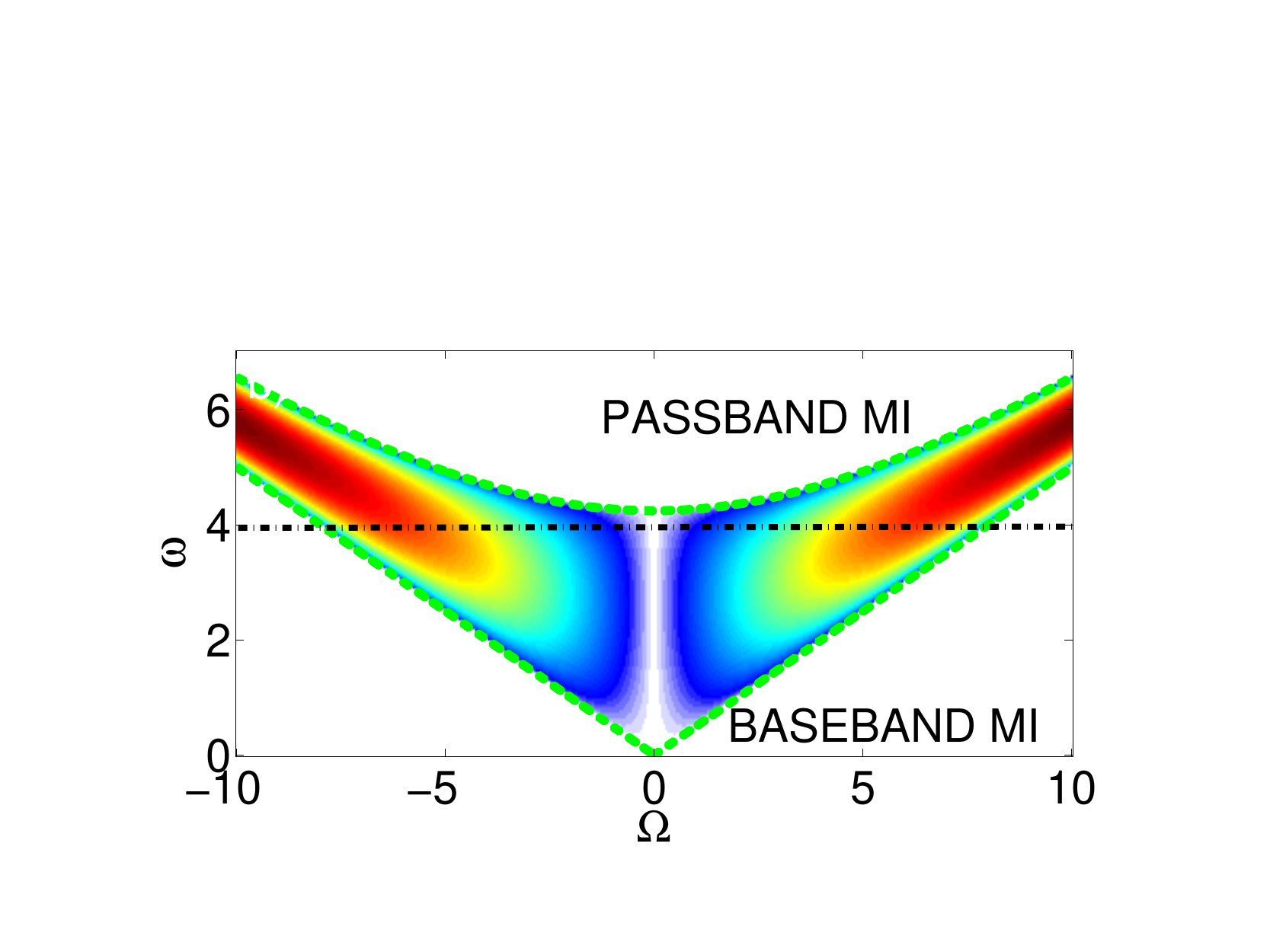}
\includegraphics[width=8cm,height=4cm]{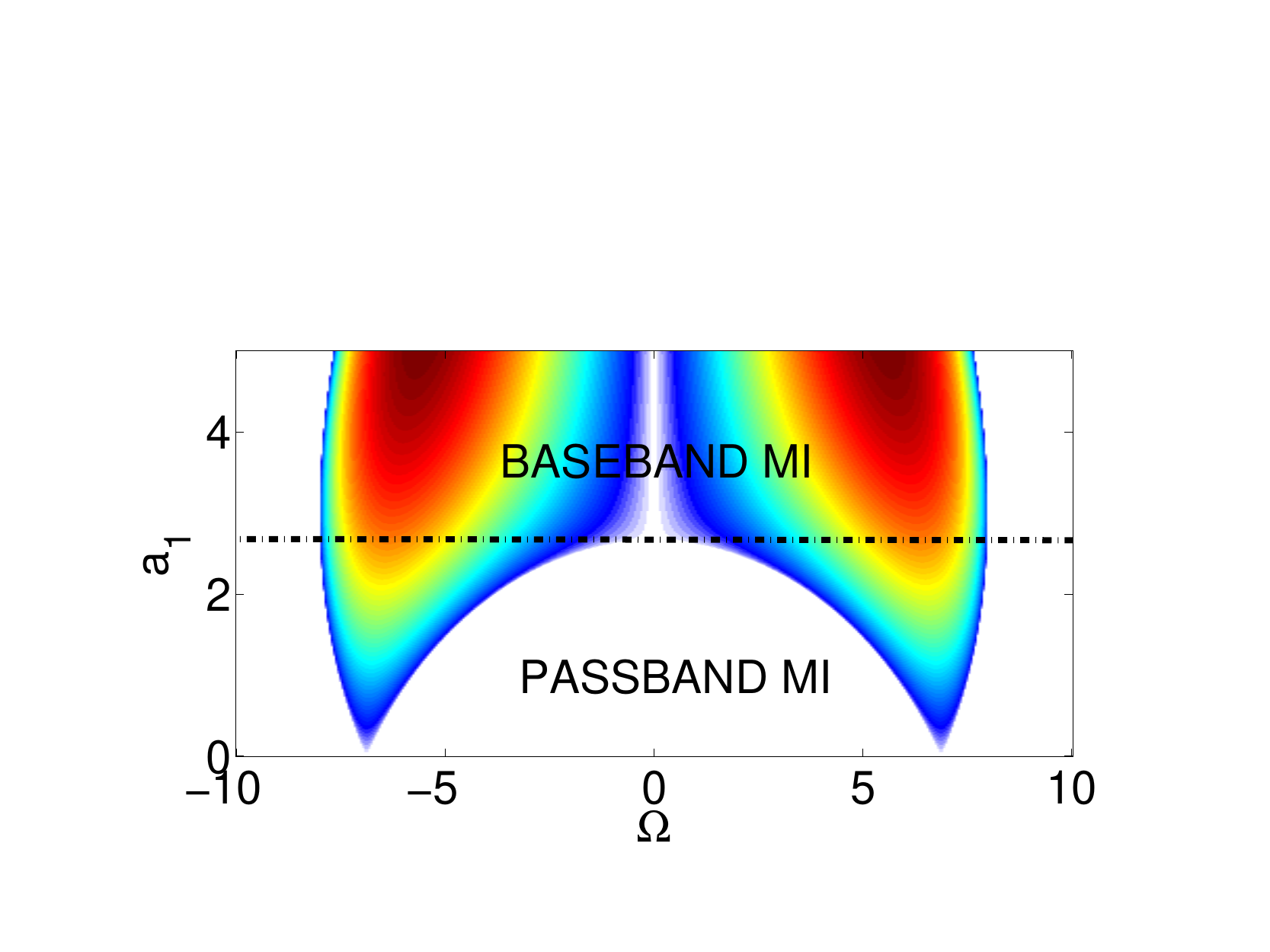}
    \end{center}
     \caption{Maps of MI gain $2G$ of the VNLSE (\ref{VNLS}).  a) MI on the $(\Omega,\omega)$
		plane, calculated for the case $a_1=3, a_2=3$, $\omega_1=-\omega_2=\omega$.  Dotted (green online) curves
		represent the analytical marginal stability condition $\Omega=2\omega$,
{$\Omega^2=\textrm{max}\{4 \omega^2 -8 a^2, 0\}$.}
		b) MI on the $(\Omega,a_1)$ plane, calculated for the case $a_2=3, \omega_1=-\omega_2=4$.
    } \label{fig_2}
\end{figure}

Figure \ref{fig_2}(a) corresponds to the case where the nonlinear background modes have opposite frequencies
($\omega_1=-\omega_2=\omega$). 
The higher $\omega$, the higher $G$. In the special case of equal background amplitudes $a_1=a_2=a$, 
the marginal stability conditions can be analytically found: {$\Omega^2=4\omega^2$, 
$\Omega^2=\textrm{max}\{4 \omega^2 -8 a^2, 0\}$. Thus, for $a^2 > \omega^2/2$ a baseband MI, which
includes frequencies that are arbitrarily close to zero, is present (i.e. $0 < \Omega^2 < 4\omega^2$). Instead, 
for $a^2 \leq \omega^2/2$, MI only occurs for frequencies within the passband range $(4\omega^2-8a^2) < \Omega^2 < 4\omega^2 $.
We may point out that the rogue waves (\ref{pere}) necessarily exist for $a^2 > \omega^2/2$.
Thus, rogue waves (\ref{pere}) and baseband MI coexist.

Figure \ref{fig_2}(b) illustrates the case of different
frequencies ($\omega_1=-\omega_2=\omega$) and input amplitudes $a_1\neq a_2$  for the nonlinear background modes.
For low values of $a_1$, only passband MI is present. By increasing $a_1$, the baseband MI condition is eventually attained.
\newline 
In order to analytically represent the condition for the occurrence of baseband MI, let us consider the limit $\Omega \rightarrow 0$. To this aim, we may rewrite the characteristic polynomial as $B(\Omega v)=\Omega^4b(v)$, and consider the polynomial $b(v)$ at $\Omega=0$, namely 
$ b(v)=v^4+b_3 v^3+ b_2 v^2+ b_1 v +b_0$, $b_0 = -16 \omega^2 (a_1^2 +  a_2^2 - \omega^2)$, $b_1 =  16q (a_1^2 - a_2^2)$, 
$b_2 = - 4 (a_1^2 + a_2^2 + 2 \omega^2)$,  $b_3 = 0$.  
Let us evaluate now the discriminant of the characteristic polynomial $B$: if the discriminant is positive, $B$ has four real roots, 
and no MI occurs. Whereas if the discriminant of $B$ is negative, there are
two real roots and two complex conjugate roots, and Eqs.(\ref{VNLS}) exhibits baseband MI.
Again, the interesting finding is that the constraint on the sign of the discriminant of the characteristic polynomial $B$, which leads to the baseband MI condition, turns out to exactly coincide with the sign constraint (\ref{eqkprima}) that is required for rogue wave existence.
Thus we may conclude that in the defocusing regime, rogue wave solutions (\ref{pere}) only exist in the subset of the parameter space where MI is present, and in particular if and only if baseband MI is present.

\begin{figure}[h]
\begin{center}
\includegraphics[width=7cm]{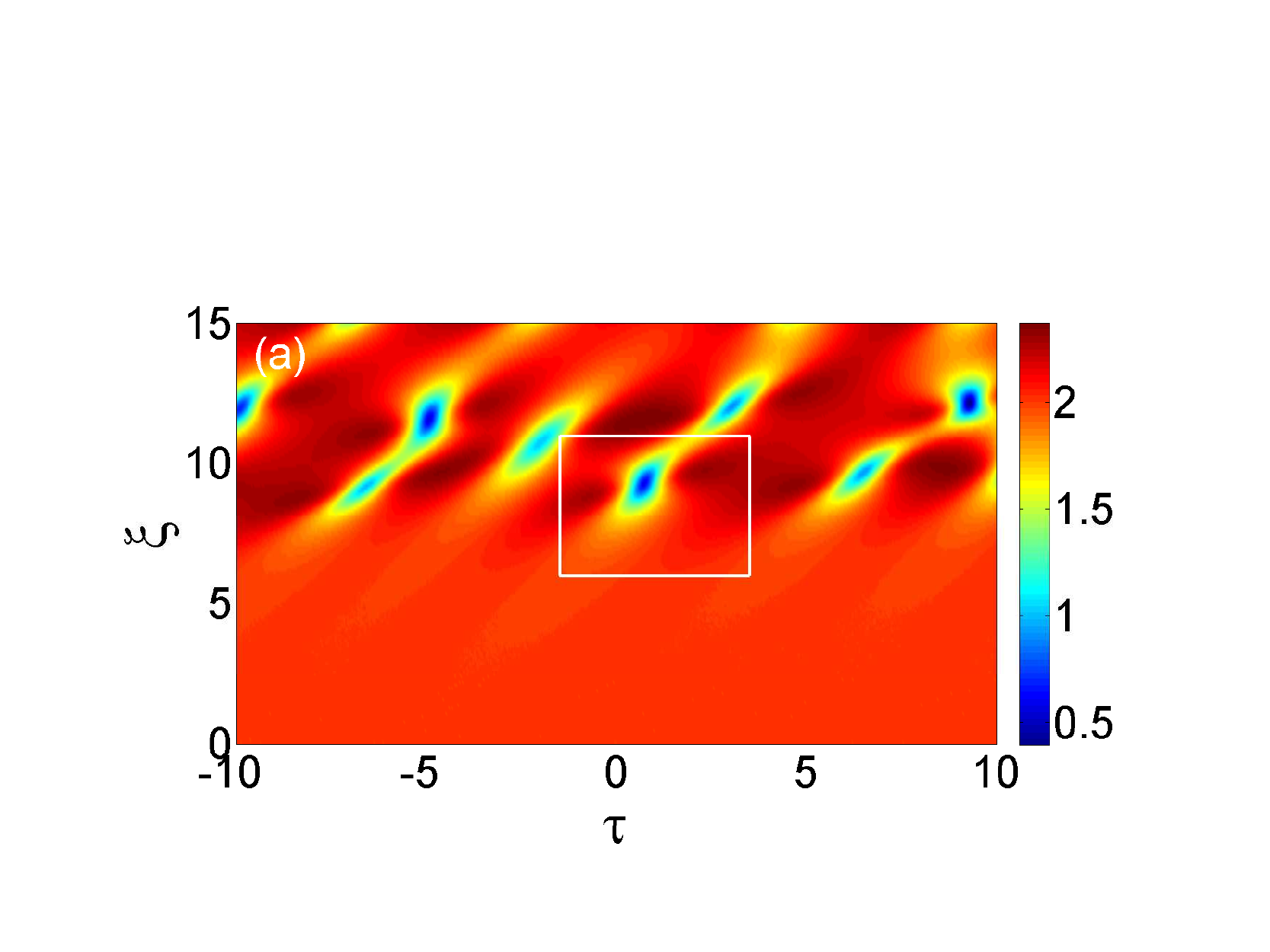}
\includegraphics[width=7cm]{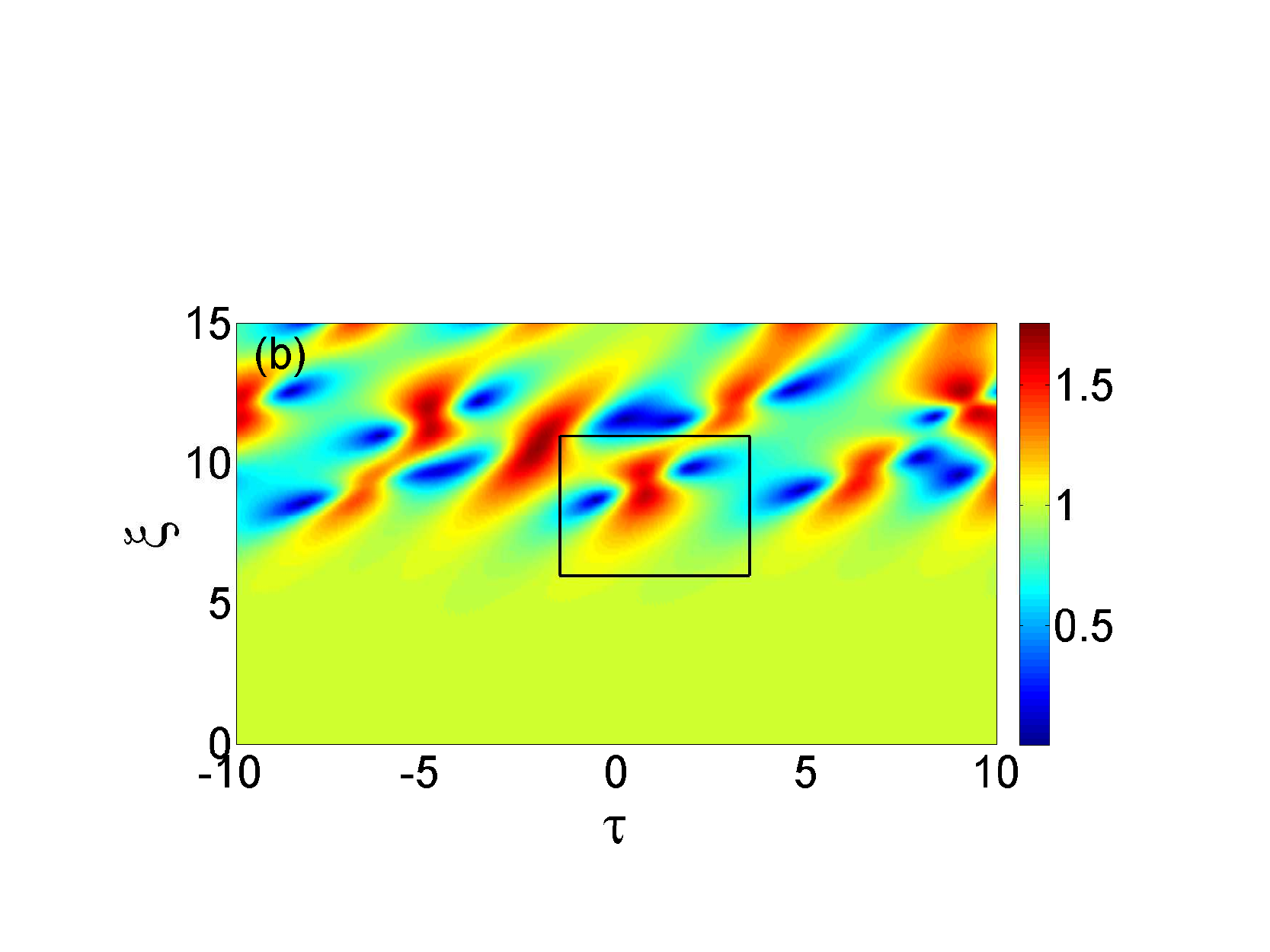}
    \end{center}
     \caption{Color plot of $|\psi^{(1)}(\tau,\xi)|$ (a) and $|\psi^{(2)}(\tau,\xi)|$
		(b) from the numerical solution of the defocusing VNLSE. The initial condition is a plane wave
		perturbed by weak random noise. Parameters: $a_1=2, a_2=1, \omega=1$. A rogue wave is highlighted by a surrounding box.
    } \label{figv4}
\end{figure}

\begin{figure}[h]
\begin{center}
\includegraphics[width=7cm]{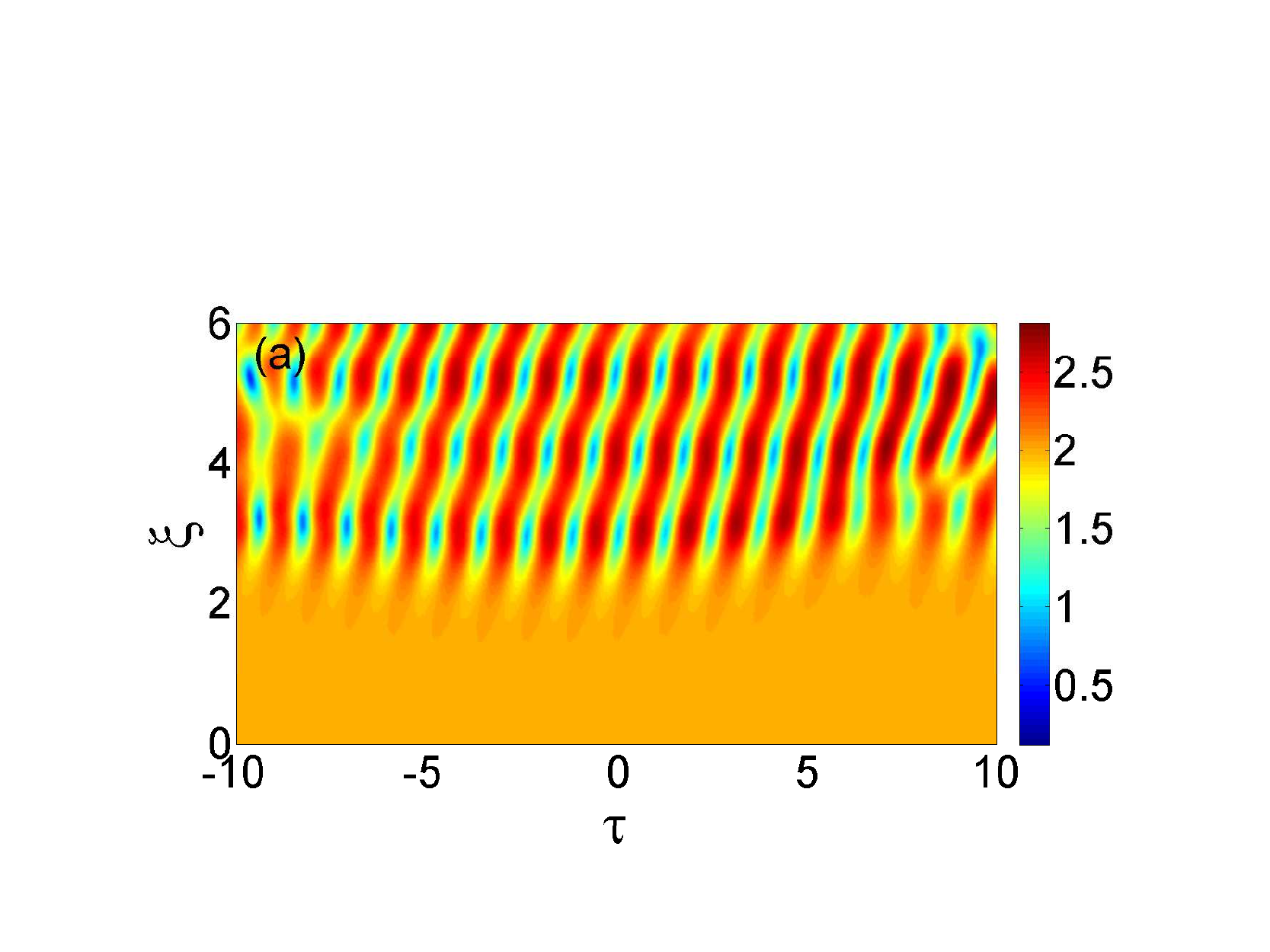}
\includegraphics[width=7cm]{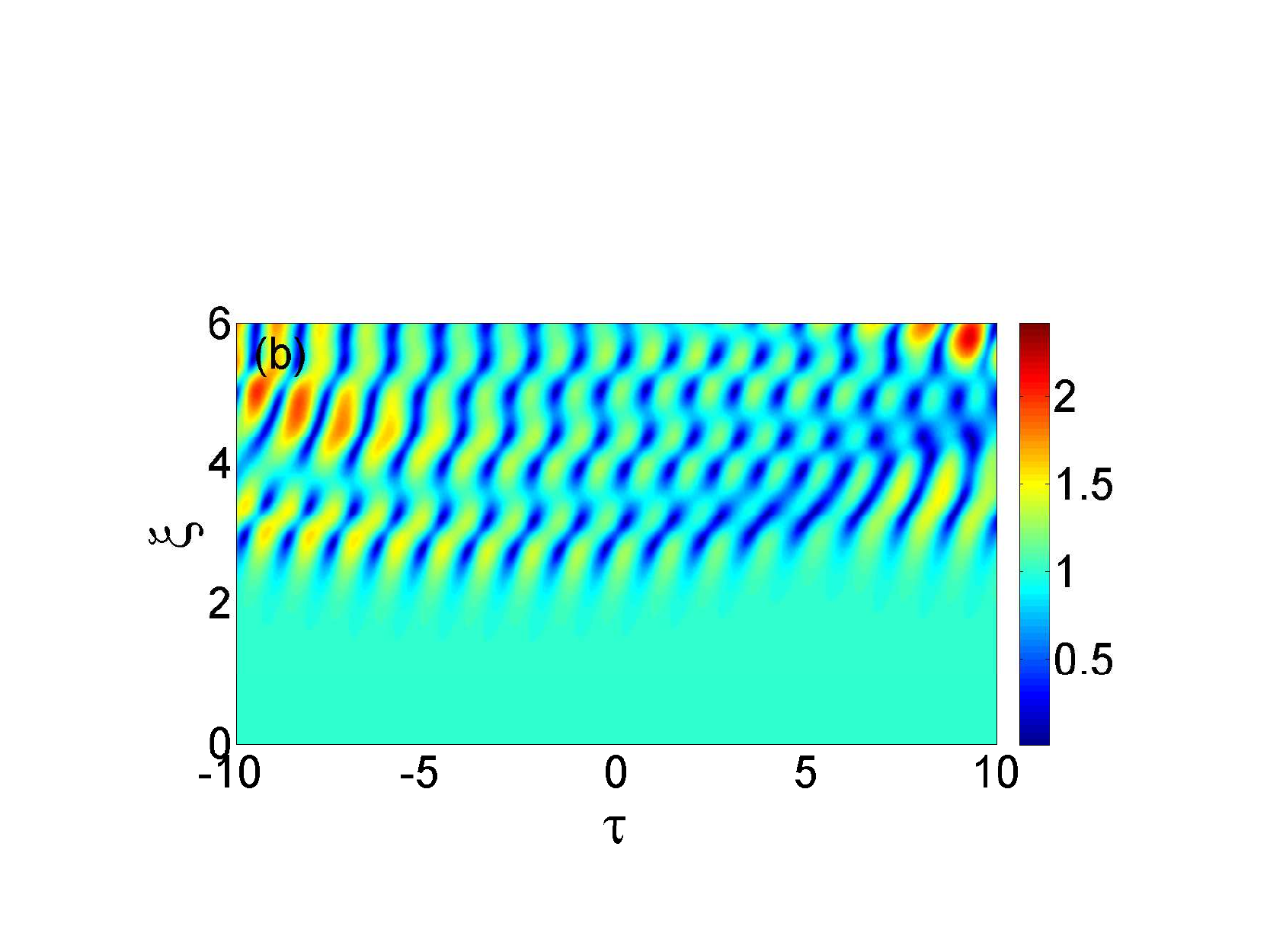}
    \end{center}
     \caption{Color plot of $|\psi^{(1)}(\tau,\xi)|$ (a) and $|\psi^{(2)}(\tau,\xi)|$
		(b) from the numerical solution of the defocusing VNLSE. The initial condition is a plane wave
		perturbed by weak random noise. Parameters: $a_1=2, a_2=1, \omega=3$. No rogue waves are generated in this case.
    } \label{figv5}
\end{figure}

Fig. \ref{figv4} and Fig. \ref{figv5} show two different numerically computed nonlinear evolutions, obtained in the case of baseband MI (leading to rogue wave generation) and 
of passband MI, respectively. These evolutions permit to highlight that the nonlinear evolution of baseband MI leads to rogue wave solutions of the VNLSE (\ref{VNLS}) . 
Figure \ref{figv4} shows the numerically computed evolution of a plane wave perturbed by a small
random noise in the baseband MI regime. After a first initial stage of linear growth of the unstable frequency modes, for $\xi>5$ the nonlinear stage of MI is reached. As we can see, MI leads to the formation of multiple isolated peaks (dips) that emerge at random positions. By carefully analyzing one of these peaks, for example the peak near the point ($\tau=0,\xi=9$), we may clearly recognize the shape of a rogue wave as it is described by the expression (\ref{pere}).   
Conversely, Fig.\ref{figv5} shows the numerically computed evolution of a plane wave perturbed by a small random noise, in the passband MI regime. After a first initial stage of linear growth of the unstable frequency modes, for $\xi>2$ the nonlinear stage of MI is reached. In this case, we may observe the generation of a train of nonlinear oscillations, with wave-numbers corresponding to the peak of MI gain ($\Omega_{max}=5$). As it was expected, no isolated peaks (dips) emerge from noise in this case,
given that the condition for the existence of rogue waves is not verified.

\section{LWSW model}
The last model we consider in our survey is the LWSW resonance. It is as well a general model that describes the interaction between a rapidly varying wave and a quasi continuous one. 
In optics the LWSW resonance rules wave propagation in negative index media \cite{tataronis08} or the optical-microvave interactions \cite{bubke03}. Whereas 
in hydrodynamics the LWSW resonance results from the interaction between capillary
and gravity waves \cite{djordjevic77}.

For our studies, we write the LWSW equations in the dimensionless form
\begin{equation}\label{LS}
\left \{ \begin{array} {lll}
i\psi^{(S)}_\xi&+& \frac{1}{2}\psi^{(S)}_{\tau \tau} + \psi^{(L)} \psi^{(S)} = 0 \\
\psi^{(L)}_\xi &-& |\psi^{(S)}|^2_{\tau} =  0, 
\end{array} \right .
\end{equation}
where $\psi^{(S)}(\xi,\tau)$ represents the short wave 
complex envelope, and $\psi^{(L)}(\xi,\tau)$ represents the long wave real field; 
$\xi$ and $\tau$ are the propagation  distance and the retarded time, 
respectively; each subscripted variable
stands for partial differentiation.

\begin{figure}[h]
\begin{center}
\includegraphics[width=7cm]{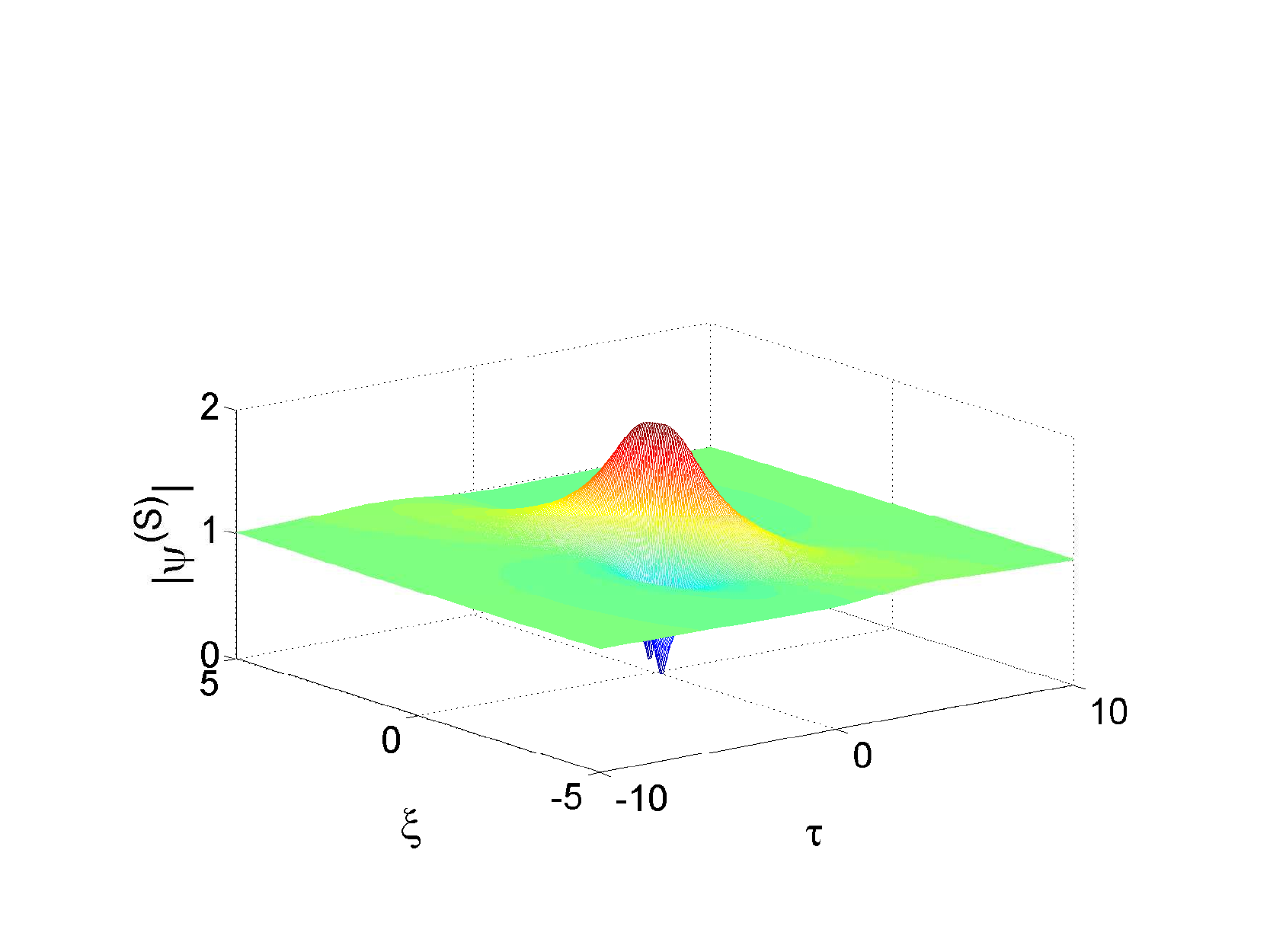}
\includegraphics[width=7cm]{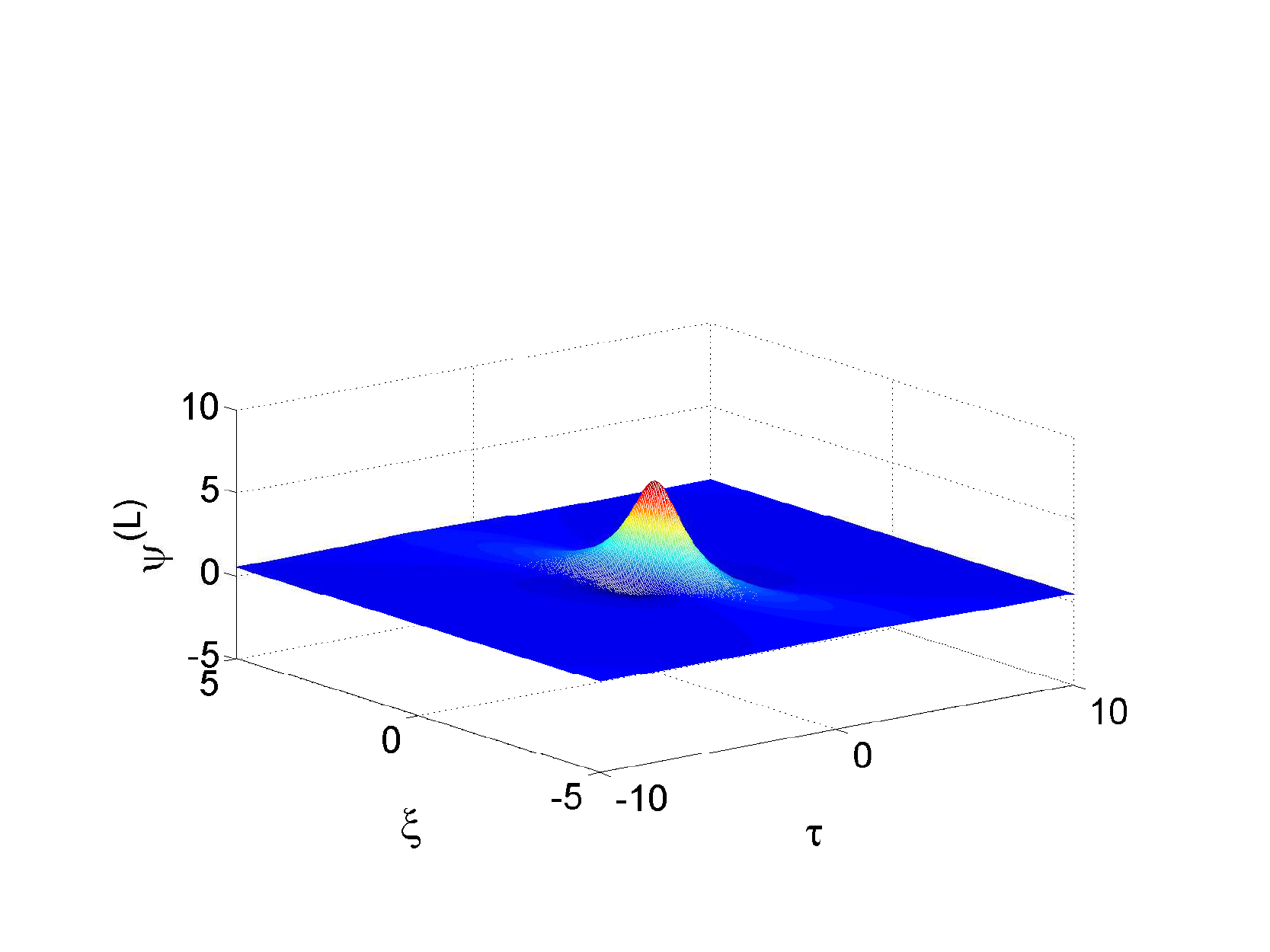}
    \end{center}
     \caption{Rogue wave envelope distributions $|\psi^{(S)}(\tau,\xi)|$
     and $|\psi^{(L)}(\tau,\xi)|$ corresponding to expressions (\ref{LS}). Here, $a=1, \omega=0, b=0.5$.
		    } \label{figLS_1}
\end{figure}

The fundamental rogue wave solution of Eqs. (\ref{LS})  has recently been reported in Ref.\cite{chen14R}, and reads as

\begin{equation}\label{LSs}
\psi^{(S)}=
  \psi_0^{(S)} \big[ 1-\frac{i \xi+ \frac{i\tau}{2m-\omega}+\frac{1}{2(2m-\omega)(m-\omega)}}{(\tau-m\xi)^2+n^2\xi^2+1/4n^2}    \big], \nonumber
\end{equation}

\begin{equation}\label{LWs}
\psi^{(L)}= b+ \frac{n^2\xi^2- (\tau-m\xi)^2+1/4n^2}{[(\tau-m\xi)^2+n^2\xi^2+1/4n^2]^2},
\end{equation}
where $\psi_0^{(S)}=a e^{i(\omega \tau - \beta \xi)}$ represents the background solution
of the short wave, defined by the amplitude $a$
($a>0$), frequency $\omega$, and wave number $\beta=\omega^2/2-b$; the amplitude $b$ ($b\geq 0$) defines
the background solution of the coupled long wave real field. The parameters
$m$ and $n$ are real, defined by $m=\frac{1}{6}[5\omega-\sqrt{3(\omega^2+l+\upsilon/l)}]$, 
$n=\pm \sqrt{(3m-\omega)(m-\omega)}$, with $\upsilon=\frac{1}{9} \omega^4+6 \omega a^2$,
$\rho=\frac{1}{2}\omega^6-\frac{1}{54}(27a^2+5\omega^3)^2$. $l=-(\rho-\sqrt{\rho^2-\upsilon^3})^{1/3}$,
for $\omega \leq -3 (2a^2)^{1/3}$, and $l=(-\rho+\sqrt{\rho^2-\upsilon^3})^{1/3}$,
for $ -3 (2a^2)^{1/3}< \omega \leq \frac{3}{2} (2a^2)^{1/3}$.
LWSW rogue waves (\ref{LWs}) depend on the real parameters $a$, $\omega$
and $b$ (see Ref. \cite{chen14R} for details 
on nonlinear wave characteristics). Figure \ref{figLS_1} shows a 
typical LWSW rogue solution.
Importantly, the existence condition for rogue waves of the LWSW model
is that $\omega\leq \frac{3}{2} (2a^2)^{1/3}$.

Let us turn our attention now to the linear stability analysis of the background solution of Eqs. (\ref{LS}).
Here a perturbed nonlinear background can be written as 
$\psi_p^{(S)}=  [a+p_S] e^{i \omega \tau-i \beta \xi}$, and $\psi_p^{(L)}= b+p_L$ 
where $p_S(\xi,\tau), p_L(\xi,\tau)$ are small complex perturbations that obey linear partial differential equations.  Whenever the perturbations $p_S, p_L$ are $\tau-$periodic with frequency $\Omega$, i.e.,
$p_S(\xi,\tau)=\eta_{s}(\xi)e^{i \Omega \tau}+\eta_{a}(\xi)e^{-i \Omega \tau}$, and recalling that 
$\psi_p^{(L)}$ is real, $p_L(\xi,\tau)=g(\xi)e^{i \Omega \tau}+g^*(\xi)e^{-i \Omega \tau}$,  the perturbation equations reduce to a $3\times3$ linear ordinary differential equation $\eta'=iM\eta$, with $\eta=[\eta_{s},\eta^*_{a},g]^T$ (here a prime stands for differentiation with respect to $\tau$). For any given real frequency $\Omega$, the generic perturbation may be expressed as a linear combination of exponentials $\exp(i w_j \xi)$ where $w_j,\;j=1,\cdots,3, $ are the three eigenvalues of the matrix:

\begin{equation}
M=
\left[\begin{array}{ccc}
 -\frac{1}{2} \Omega^2-\omega \Omega & 0 & a\\
0 & \frac{1}{2} \Omega^2-\omega \Omega & -a\\
\Omega a & \Omega a &0
\end{array}\right].
\end{equation}

Since the entries of the matrix $M$ are all real, the eigenvalues $w_j$ are either real, or they appear as complex conjugate pairs. These eigenvalues are obtained as the roots of the characteristic polynomial $B(w)$ of the matrix $M$:

\begin{align}
B(w)&=B_3 w^3+ B_2 w^2+ B_1 w +B_0,\\
\nonumber B_0 &= a^2 \Omega^3,\;\; B_1 = \omega^2 \Omega^2 -\Omega^4/4,\;\;B_2 =  2 \omega \Omega,\;\; B_3 = 1. 
\end{align}

MI occurs whenever $M$ has an eigenvalue $w$ with a negative imaginary part, i.e., $\textrm{Im}\{w\} < 0$. 
Indeed, if the explosive rate is $G(\Omega) = -\textrm{Im}\{w\} >0$, perturbations  
grow larger exponentially like $\exp(G \xi)$ at the expense of the pump waves. 
By calculating the discriminant of the polynomial $B$, one finds
$\Delta= \Omega^6 (\frac{1}{16}\Omega^6-\frac{1}{2}\Omega^4-\omega (9a^2- \omega^3)\Omega^2+4a^2 \omega^3-27a^4)$.
If the discriminant $\Delta$ is positive, the polynomial $B$ has real roots, 
and no MI occurs. Conversely if the discriminant $\Delta$ is negative, the polynomial $B$ has two
complex conjugate roots, which means that MI is present for Eqs.(\ref{LS}). The marginal stability
curves, corresponding to $\Delta=0$, can thus be calculated.
Figure \ref{figLS_stab}  shows a typical  MI gain spectrum of the LWSW Eqs. (\ref{LS}):
as one can see, there exist regions of either baseband or passband MI.

\begin{figure}[h]
\begin{center}
\includegraphics[width=8cm]{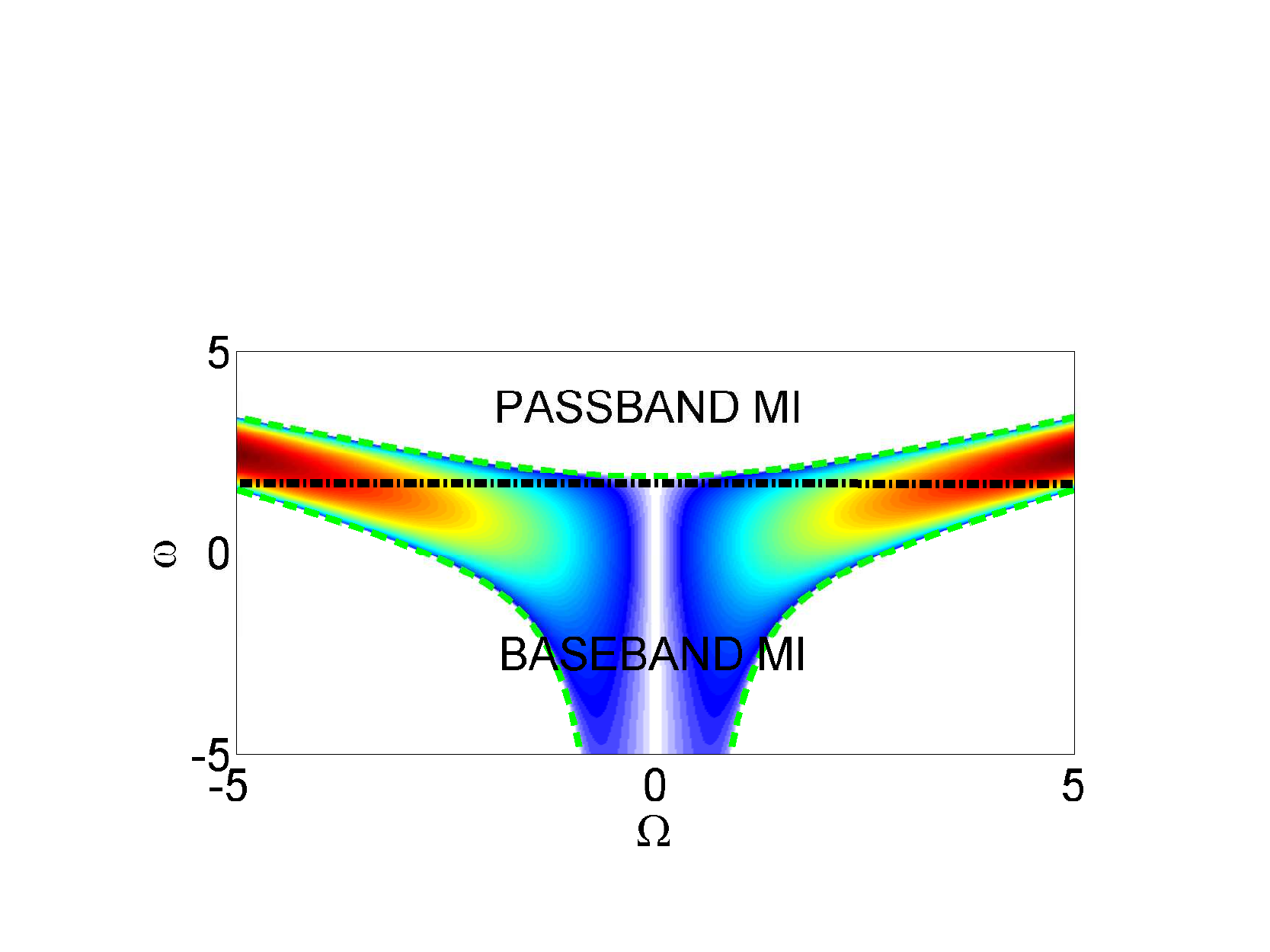}
    \end{center}
     \caption{Maps of MI gain $2G$ of the LWSW Eqs. (\ref{LS}). MI on the $(\Omega,\omega)$
		plane, calculated for the case $a=1$, Dashed (green online) curves
		represent the analytical marginal stability condition 
		$\Omega^6 (\frac{1}{16}\Omega^6-\frac{1}{2}\Omega^4-\omega (9a^2- \omega^3)\Omega^2+4a^2 \omega^3-27a^2)=0$.
		 } \label{figLS_stab}
\end{figure}

\begin{figure}[htb]
\begin{center}
\includegraphics[width=8cm]{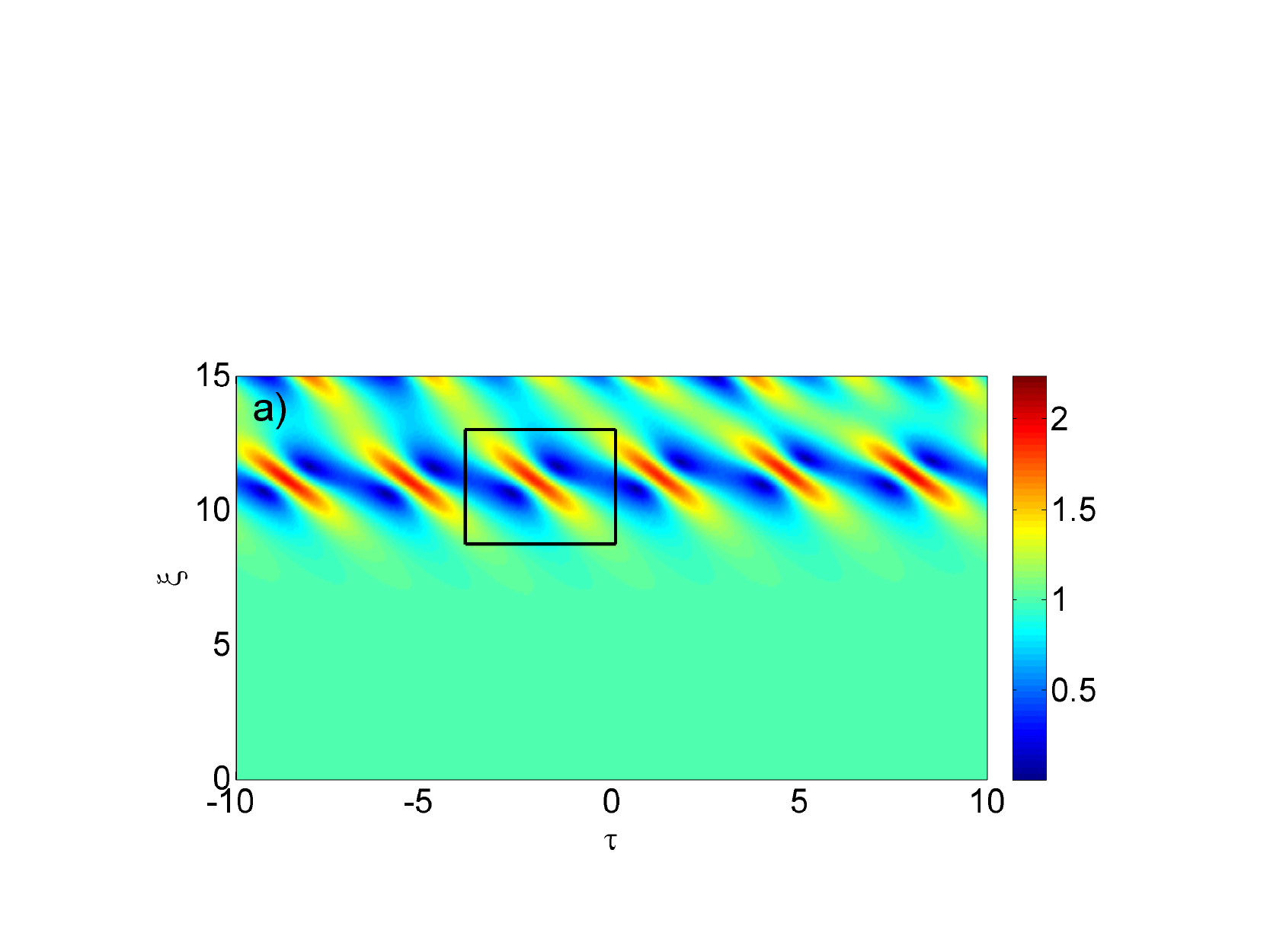}
\includegraphics[width=8cm]{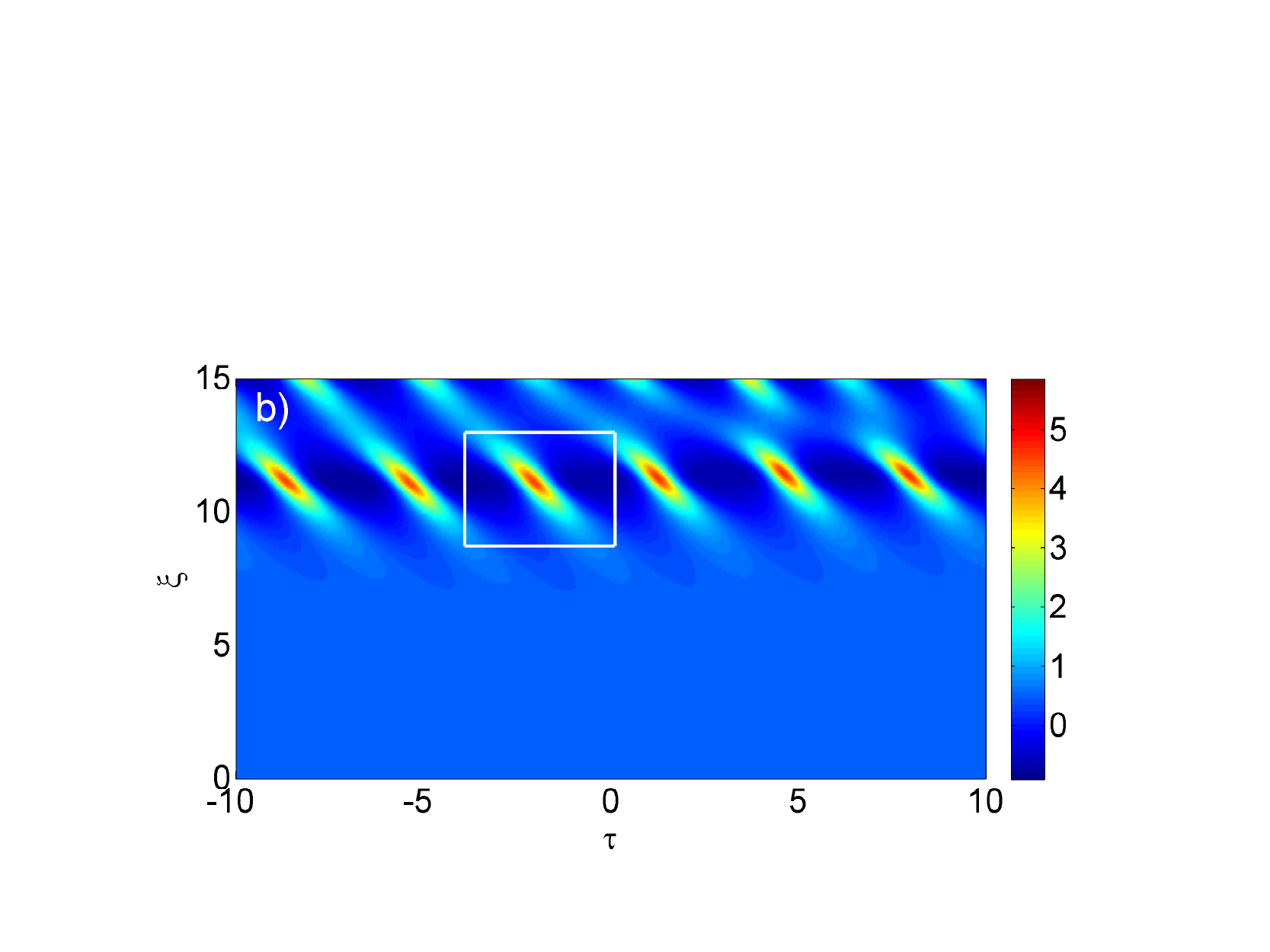}
    \end{center}
     \caption{Color plot of $|\psi^{(S)}(\tau,\xi)|$ (a) and $|\psi^{(L)}(\tau,\xi)|$
		(b) from the numerical solution of the LWSW equation. The initial condition is a plane wave		perturbed by weak random noise. Parameters: $a=1, B=0.5, \omega=0$. A rogue wave is highlighted by a surrounding box.} \label{figLS_rogue}
\end{figure}

\begin{figure}[htb]
\begin{center}
\includegraphics[width=8cm]{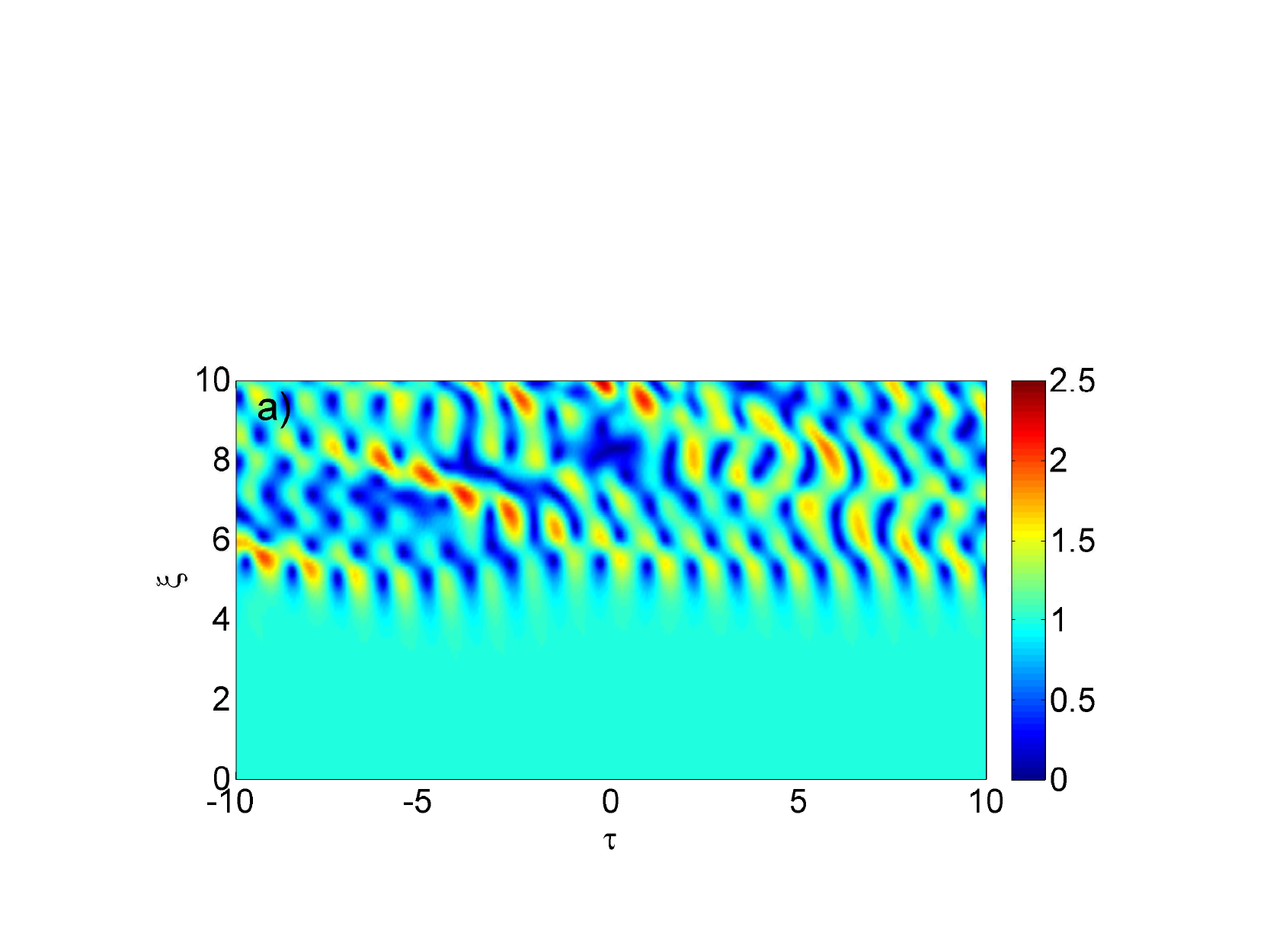}
\includegraphics[width=8cm]{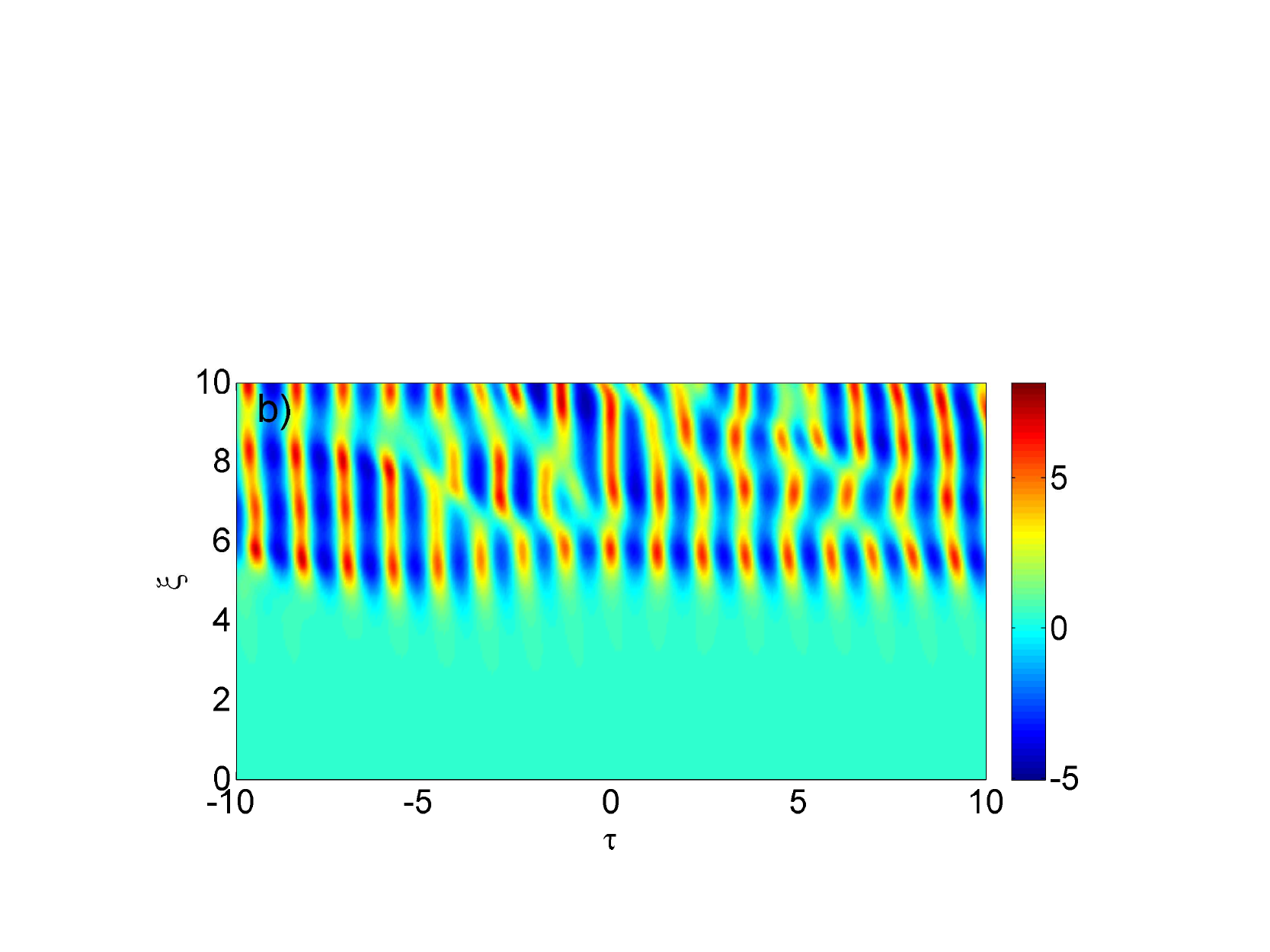}
    \end{center}
     \caption{Color plot of $|\psi^{(S)}(\tau,\xi)|$ (a) and $|\psi^{(L)}(\tau,\xi)|$
		(b) from the numerical solution of the LWSW equation. The initial condition is a plane wave		perturbed by weak random noise. Parameters: $a=1, B=0.5, \omega=2.5$. No rogue wave is generated in this case.} \label{figLS_rogue}
\end{figure}

As in previous sections, let us proceed now to discuss the MI behavior in the limit situation where $\Omega \rightarrow 0$, a condition which characterizes the occurrence of baseband MI. In this regime, the discriminant of the polynomial $B$ reduces to $\Delta=4a^2\omega^3-27a^4$, which leads to the MI condition $\omega < \frac{3}{2} (2a^2)^{1/3}$. Again, the baseband MI condition turns out to exactly coincide with the condition for the existence of rogue wave solutions of Eqs. (\ref{LS}).

Figure \ref{figLS_rogue} shows a numerical solution of LWSW, obtained in the case of baseband MI (leading to rogue wave generation), showing the evolution of a plane wave perturbed by a small
random noise. After a first initial stage of linear growth of the unstable frequency modes, for $\xi>8$ the nonlinear stage of MI is reached. As we can see, MI leads to the formation of multiple isolated peaks that emerge at random positions. By carefully analyzing one of these peaks, we may clearly recognize the shape of a rogue wave as it is described by the expression (\ref{LSs}).   

\section{Conclusions}
In this work we studied the existence and the properties of rogue wave solutions 
in different integrable nonlinear wave evolution models which are of widespread use both in optics and in hydrodynamics. Namely, we considered 
the  Fokas-Lenells equation, 
the defocusing vector nolinear Schr\"odinger equation and the 
long-wave-short-wave resonance.
{\bf We found out that in all of these models rogue waves, which can be modeled as rational solutions, only exist in the subset of parameters where MI is present, but if and only if the MI gain band also contains the zero-frequency perturbation as a limiting case (baseband MI).}
 %We found out thatin all of these models rogue wave solutions only exist in the subset of parameters where MI 
%is present, but if and only if the MI gain band also contains the 
%zero-frequency perturbation as a limiting case (baseband MI). 
We have numerically confirmed that in the baseband-MI regime 
rogue waves can indeed be excited from a noisy input cw background. Otherwise, when there is passband MI we only observed the generation of nonlinear wave oscillations.  
Based on the above findings, we are led to believe that the conditions for simultaneous rogue wave existence and of baseband MI 
may also be extended to other relevant and integrable and non-integrable
physical models of great interest for applications, for instance consider frequency conversion models \cite{baro10,confo12}
where extreme wave events and complex breaking beaviours are known to 
place \cite{confo06,confo13}.

\section*{Acknowledgments}
The present research was supported by the Italian
Ministry of University and Research 
(MIUR, Project No. 2012BFNWZ2), by the Agence 
Nationale de la Recherche (projects TOPWAVE and NoAWE).


\begin{thebibliography}{99}
%%Do not include separate BibTeX files; if BibTeX is used,
%% paste the output (contents of .bbl file) here.

\bibitem{hopkin04},
%M. Hopkin, ``Sea snapshots will map frequency of freak waves,'' 
Nature \textbf{430}, 492 (2004). 

\bibitem{perkin06}
S. Perkins, %``Dashing Rogues: freak ocean waves pose threat to ships, deep-sea oil platforms,'' 
Science News \textbf{170}, 328 (2006).  


\bibitem{pelinosky08}
E. Pelinovsky and C. Kharif, \textit{Extreme Ocean Waves} (Springer, Berlin, 2008). 

\bibitem{kharif09}
C. Kharif, E. Pelinovsky, and A. Slunyaev, \textit{Rogue Waves in the Ocean} (Springer, Heidelberg, 2009). 


\bibitem{dud14}
J. M. Dudley, F. Dias, M. Erkintalo, and G. Genty,
%``Instabilities, breathers and rogue waves in optics,'' 
Nat. Photon. \textbf{8}, 755 (2014). 


%\bibitem{akhmediev2010sp}
%N. Akhmediev and E. Pelinovsky, %\textit{Editorial-Rogue Waves-Towards a Unifyng Concept?: Discussions and Debates}, 
%Eur. Phys. J. Special Topics \textbf{185}, 1 (2010).


\bibitem{onorato13}
M. Onorato, S. Residori, U. Bortolozzo, A. Montina, and F.T. Arecchi, 
%``Rogue waves and their generating mechanisms in different physical contexts,'' 
Phys. Rep. \textbf{528}, 47 (2013).


\bibitem{peregrine83}
D.H. Peregrine, 
%``Water waves, nonlinear Schr\"odinger equations and their solutions,''
J. Australian Math. Soc. Ser. B \textbf{25}, 16 (1983). 


\bibitem{kibler10}
B. Kibler, J. Fatome, C. Finot, G. Millot, F. Dias, G. Genty, N. Akhmediev, and J.M.
Dudley, 
%``The Peregrine soliton in nonlinear fibre optics,''
Nat. Phys. \textbf{6}, 790 (2010). 
%

\bibitem{amin11}
A. Chabchoub, N.P. Hoffmann, and N. Akhmediev, 
%``Rogue waves in a water wave tank,''
Phys. Rev. Lett. \textbf{106}, 204502 (2011).

\bibitem{bailung11}
H. Bailung, S.K. Sharma, and Y. Nakamura, 
%``Observation of Peregrine solitons in a multicomponent plasma with negative ions,''
Phys. Rev. Lett. \textbf{107}, 255005 (2011).


\bibitem{lecap12}
C. Lecaplain, Ph. Grelu, J.M. Soto-Crespo, and N. Akhmediev,
%``Super rogue waves: observation of a higher-order breather in water waves,''
Phys. Rev. Lett. \textbf{108}, 233901 (2012). 


\bibitem{ankiew10}
A. Ankiewicz, J.M. Soto-Crespo and N. Akhmediev,
%``Rogue waves and rational solutions of the Hirota equation,''
Phys. Rev. E \textbf{81}, 046602 (2010).


\bibitem{bandel12}
U. Bandelow and N. Akhmediev,
%``Sasa-Satsuma equation: soliton on a background and its limiting cases,''
Phys. Rev. E \textbf{86}, 026606 (2012).

\bibitem{chen14}
S. Chen and L. Y. Song
%``Peregrine solitons and algebraic soliton pairs in Kerr media considering space-time correction,''
Phys. Lett. A \textbf{378}, 1228 (2014). 

\bibitem{baronio12}
F. Baronio, A. Degasperis, M. Conforti, and S. Wabnitz, 
%``Solutions of the vector nonlinear Schr\"odinger equations: evidence for deterministic rogue waves,''
Phys. Rev. Lett. \textbf{109}, 044102 (2012).

\bibitem{liu2013}
L.C. Zhao and J. Liu,
%``Rogue-wave solutions of a three-component coupled nonlinear Schr\"odinger equation,''
Phys. Rev. E \textbf{87}, 013201 (2013).

\bibitem{zhai2013}
B.G. Zhai, W.G. Zhang, X.L. Wang, H.Q. Zhang,
%``Multi-rogue waves and rational solutions of the coupled nonlinear
Schr\"odinger equations,''
Nonlinear Anal-Real \textbf{14}, 14-27 (2013).

\bibitem{baronio14}
F. Baronio, M. Conforti, A. Degasperis, S. Lombardo, M. Onorato, and S. Wabnitz, 
%``Vector rogue waves and baseband modulation instability in the defocusing regime,''
Phys. Rev. Lett. \textbf{113}, 034101 (2014).

\bibitem{baronio13}
F. Baronio, M. Conforti, A. Degasperis, and S. Lombardo, 
%``Rogue waves emerging from the resonant interaction of three waves,''
Phys. Rev. Lett. \textbf{111}, 114101 (2013).


\bibitem{chen13}
S. Chen and L. Y. Song,
%``Rogue waves in coupled Hirota systems,''
Phys. Rev. E  \textbf{87}, 032910 (2013). 



\bibitem{chen14R}
S. Chen, Ph. Grelu, and J.M. Soto-Crespo,
%``Dark- and bright-rogue-wave solutions for media with long-wave short-wave resonance,''
Phys. Rev. E  \textbf{89}, 011201(R) (2014). 

\bibitem{zakh09}
V. E. Zakharov and L. A. Ostrovsky,
%``Modulation instability: The beginning,'' 
Physica D  \textbf{238}, 540 (2009). 


\bibitem{ruderman10}
M.S. Ruderman, 
%``Freak waves in laboratory and space plasmas,''
Eur. Phys. J. Special Topics \textbf{185}, 57 (2010).

\bibitem{sluniaev10}
A. Sluniaev, 
%``Freak wave events and the wave phase coherence,''
Eur. Phys. J. Special Topics \textbf{185}, 67 (2010).

\bibitem{kharif10}
C. Kharif and J. Touboul,
%``Under which conditions the Benjamin-Feir instability may spawn an extreme wave event: A fully nonlinear approach,''
Eur. Phys. J. Special Topics \textbf{185}, 159 (2010).


\bibitem{fok95}
A. S. Fokas,
%``On a class of physically important integrable equations,''
Physica D \textbf{87}, 145 (1995).

\bibitem{len09}
J. Lenells, %\textit{}, 
%``Exactly solvable model for nonlinear pulse propagation in optical fibers,''
Stud. Appl. Math. \textbf{123}, 215 (2009).

%
%\bibitem{shen12}
%Y. Shen, N. Whitaker, P.G. Kevrekidis, N.L. Tsitsas, and D.J. Frantzeskakis, 
%``Ultrashort pulses and short-pulse equations in 2+1 dimensions,''
%Phys. Rev. A \textbf{86}, 023841 (2012).

\bibitem{miguel_crossing}
M. Onorato, A. R. Osborne, and M. Serio,
%``Modulational Instability in Crossing Sea State: A possible Mechanism for the formation of Freak Waves,''
Phys. Rev. Lett. {\bf 96}, 014503 (2006).

\bibitem{meniuk}
P. K. A. Wai and C. R. Menyuk,
%``Polarization Mode Dispersion, Decorrelation, and Diffusion in Optical Fibers with Randomly Varying Birefringence,''
J. Lightwave Technol. {\bf 14}, 148 (1996).

\bibitem{segev}
Z. Chen, M. Segev, T. H. Coskun, D. N. Christodoulides, and Y. S. Kivshiar,
%``Coupled photorefractive spatial-soliton pairs,''
J. Opt. Soc. Am. B {\bf 11}, 3066 (1997).

\bibitem{tataronis08}
A. Chowdhury and J. A. Tataronis,
%``Long Wave–Short Wave Resonance in Nonlinear Negative Refractive Index Media,''
Phys. Rev. Lett. {\bf 100}, 153905 (2008).

\bibitem{bubke03}
K. Bubke, D. C. Hutchings, U. Peshel, and F. Lederer,
%``Dynamics and stability of solitary waves in optical-microwave interaction,''
Phys. Rev. E {\bf 67}, 016611 (2003).

\bibitem{djordjevic77}
V. D. Djordjevic and L. G. Redekopp,
%``On two-dimensional packets of capillary-gravity waves,''
J. Fluid Mech. {\bf 79}, 703 (1977).


%\bibitem{len09bis}
%J. Lenells and A. S. Fokas %\textit{}, 
%Nonlinearity \textbf{22}, 11 (2009).




\bibitem{baro10}
F. Baronio, M. Conforti, C. De Angelis, A. Degasperis, M. Andreana, V. Couderc, and A. Barthelemy,
%``Velocity-locked solitary waves in quadratic media,''
Phys. Rev. Lett. \textbf{104}, 113902  (2010).


\bibitem{confo12}
M. Conforti, F. Baronio, and S. Trillo,
%``Dispersive shock waves in phase-mismatched second-harmonic generation,''
Opt. Lett. \textbf{37}, 1082-1084 (2012).


\bibitem{confo06}
M. Conforti, F. Baronio, A. Degasperis, and S. Wabnitz,
%``Inelastic scattering and interactions of three-wave parametric solitons,''
Phys. Rev. E \textbf{74}, 065602  (2006).


\bibitem{confo13}
M. Conforti, F. Baronio, and S. Trillo,
%``Competing wave-breaking mechanisms in quadratic media,''
Opt. Lett. \textbf{38}, 1648 (2013).
%
%``,'' 

\end{thebibliography}
\end{document}